\documentclass[]{aa} 
\usepackage[varg]{txfonts}
\usepackage[switch]{lineno}

\usepackage{amsmath,amssymb,mathrsfs,graphicx,float,latexsym,url}
\usepackage{epstopdf,ragged2e}
\usepackage[linktocpage]{hyperref}
\makeatletter
\renewcommand*\aa@pageof{, page \thepage{} of \pageref*{LastPage}}
\makeatother

\usepackage[usenames,dvipsnames]{color}
\definecolor{darkred}{rgb}{0.5,0,0}
\definecolor{darkgreen}{rgb}{0,0.5,0}
\definecolor{darkblue}{rgb}{0,0,0.5}
\definecolor{prussian}{rgb}{0.0, 0.19, 0.33}
\definecolor{richelectricblue}{rgb}{0.03, 0.57, 0.82}
\definecolor{teal}{rgb}{0.0, 0.5, 0.5}
\definecolor{mediumseagreen}{rgb}{0.24, 0.7, 0.44}
\definecolor{lust}{rgb}{0.9, 0.13, 0.13}
\definecolor{ballblue}{rgb}{0.13, 0.67, 0.8}
\definecolor{darkcyan}{rgb}{0.0, 0.55, 0.55}
\definecolor{mountainmeadow}{rgb}{0.19, 0.73, 0.56}
\definecolor{palecarmine}{rgb}{0.69, 0.25, 0.21}
\definecolor{richcarmine}{rgb}{0.84, 0.0, 0.25}
\definecolor{tangelo}{rgb}{0.98, 0.3, 0.0}
\definecolor{venetian}{rgb}{0.784,0.031,0.082}
\definecolor{bdfrance}{rgb}{0.192,0.549,0.906}
\usepackage{mathtools}
\usepackage{amsbsy}
\usepackage{bm}
\usepackage{float}

\hypersetup{colorlinks=true, citecolor=blue,
linkcolor=blue, urlcolor=magenta}

\newcommand{\hatx}{\hat{X}_{ij}}
\newcommand{\rhoD}{\rho_{\Delta}}
\newcommand{\rhoR}{\rho_{\rm R}}
\newcommand{\VD}{V_{\Delta}}
\newcommand{\VR}{V_{\rm R}}
\newcommand{\nD}{\nu_{\Delta}}
\newcommand{\nR}{\nu_{\rm R}}

\newcommand{\bea}{\begin{eqnarray}}
\newcommand{\eea}{\end{eqnarray}}
\def\apjs{{The Astrophysical Journal Supplement}}
\def\apjl{{ApJL}}

\def\aap{{A\&A}}

\def\04a{{2004 a}}
\def\04b{{2004 b}}

\begin{document}

\title{Torque lags from magnetic unlocking in neutron stars and gravitational-wave smearing}

\date{\today}

\author{Arthur G. Suvorov \inst{1} and Armen Sedrakian \inst{2,3}}
\institute{\inst{1} Theoretical Astrophysics, IAAT, University of
  T{\"u}bingen, T{\"u}bingen, D-72076, Germany \\
  \inst{2} {Institute of Theoretical Physics, University of Wroc\l{}aw, 50-204 Wroc\l{}aw, Poland} \\
\inst{3} {Frankfurt Institute for Advanced Studies,  60438 Frankfurt am Main, Germany} \\
\email{a.suvorov@uni-tuebingen.de}; \email{armen.sedrakian@uwr.edu.pl}}

\date{Received ??; ??}

\abstract{In canonical twisted-torus models of hydromagnetic equilibrium in neutron stars, the toroidal field is confined to regions where the poloidal field lines are closed. As spindown torques applied at the stellar surface need not be communicated efficiently to such magnetically unlocked regions, toroidal zones or magnetic voids can rotate faster than the electromagnetically braked part of the star. Continuous gravitational radiation from the associated non-corotating quadrupole can therefore be emitted at higher frequencies than poloidal ones to a degree that depends on the diffusive microphysics controlling angular momentum transport. Dissipative processes can suppress long-lived lags, and we estimate the relevant timescales and their sensitivity to microphysics. Nevertheless, we show that a quadrupole component whose pattern speed differs from the observed spin is shifted, or smeared, over a finite frequency band. Single-template or narrowband upper limits therefore constrain only the quadrupole contribution lying inside the searched frequency window, not necessarily the full frequency-integrated quadrupole. Using simple models, we quantify this spectral redistribution over finite observation windows and discuss its implications for continuous-wave searches.}

     \titlerunning{Torque lags and spectral smearing}
   \authorrunning{A. G. Suvorov, A. Sedrakian}
\maketitle

\section{Introduction} \label{sec:intro}

Multimessenger data from the landmark gravitational-wave (GW) event
GW170817 have proven invaluable in constraining both micro-
\citep{rip23,suv24} and macrophysical \citep{baus17,LIGOScientific18}
elements of neutron star interiors.  Another type of signal that is
hoped to be observed soon involves {continuous GW emission}~\citep{jan98,abb08,wette23}, as it holds promise for
additional information that is inaccessible electromagnetically.  For
instance, hydromagnetic deformations produce a mass-quadrupole moment
that varies in time as the star spins, leading to persistent radiation
that encodes the magnetic geometry \citep{c53,fr12,lm13}.  With the
exception of precession \cite[e.g.,][]{jones02,gao20,suvkok21a},
estimates for continuous strains from magnetic deformations have
thus far been explored in the context of rigid rotation: if the whole
body spins at a linear frequency of $\nR$, the dominant spectral feature resides at $2 \nR$, which simplifies the analysis significantly.

Within minutes after birth, a solid crust forms and then grows as crystallisation proceeds through
the outer layers as the neutron star cools.  The magnetic field anchored there induces a torque
which acts at the surface, leading to gradual spindown.  However, causality
demands that there is a delay, related to the propagation speed of
magnetohydrodynamic (MHD) modes or otherwise, before the core can
respond \citep{easson79}.  The theory behind torque lags has been
studied extensively \citep{link91,mel12} and has important
implications for GW emission: if some regions of the star rotate
differentially then the overall spectrum may not consist of a single
peak but rather could be {smeared}.

\begin{figure}
\includegraphics[width=0.46\textwidth]{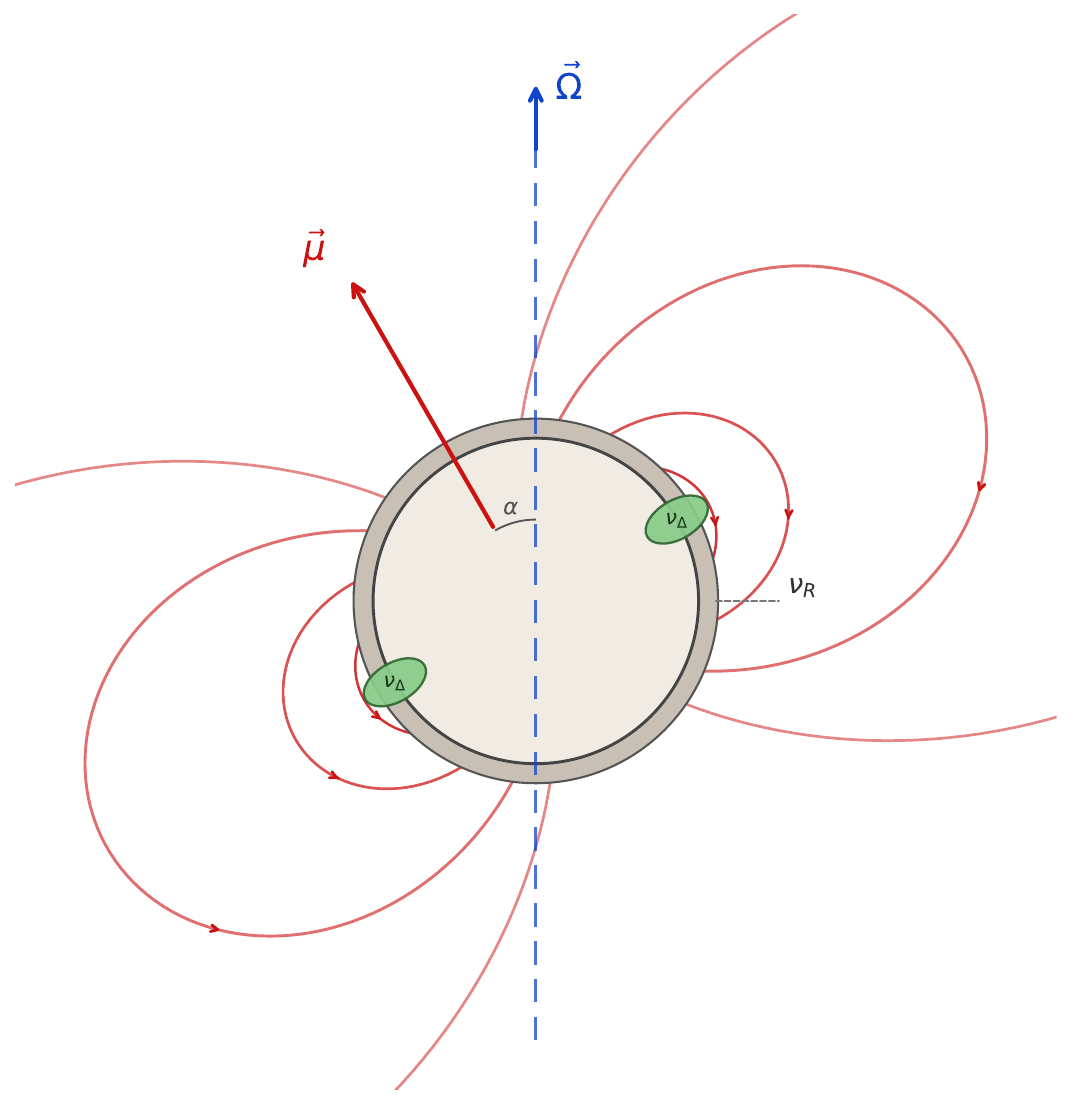}
\caption{Sketch of a twisted-torus equilibrium in a neutron star.  A
  torque is applied at the crust (shaded) which diffuses quickly there
  and through poloidally-threaded regions (lines), enforcing
  corotation ($\nR$).  In magnetically-unlocked regions (equatorial
  circles), angular momentum instead diffuses much more slowly,
  leading to faster rotation ($\nD$).
\label{fig:schematic}}
\end{figure}

Both analytic \citep{glamp15} and numerical \citep{anz20} investigations
indicate that the magnetic field plays a crucial r{\^o}le in relieving
torque lags.  This is because the field is frozen into the fluid due
to the conductivity of neutron-star matter \citep{baym69}, so that
angular momentum is transported on the (fast) Alfv{\'e}n timescale.
On the other hand, the prevailing `twisted torus' equilibrium model
for the interior consists of a threading poloidal field with a
stabilising toroidal field that is confined within the neutral curves
\citep{bra06}.  Meissner expulsions in superconducting regions could
also lead to disconnected, and possibly quite complicated magnetic topologies 
\citep{lan25,Das2025,Das2026,sed26}.

Core regions encased by closed field lines may be weakly magnetically
coupled to the crust \citep{mel12}; angular-momentum transport there
can then be governed by slower viscous, superfluid, or
composition-limited channels.  Given that toroidal fields could be
large in some or all stars, as anticipated from magnetothermal
\citep{pv19,pons11} and stability \citep{akg13,herb17} considerations,
the dominant quadrupole moment may originate from zones that rotate
differentially.  A sketch is shown in Figure~\ref{fig:schematic}: note
the importance of the misalignment between the magnetic
$(\boldsymbol{\mu}$) and rotation $(\boldsymbol{\Omega})$ axes in
order for deformations to be non-axisymmetric and instigate GW
radiation.

In single-template searches, where phase-locking between
electromagnetic and GW signals is assumed, any smearing will be missed
and thus `toroidal quadrupoles' could escape detection.  In narrowband
searches, by contrast, a typical mismatch of order
$\delta = \Delta \nu / \nR \sim 10^{-3}$ is permitted to
account for possible mismatches between the electromagnetic and GW-emitting components \citep{Aasi2015Narrowband,Abbott2017Narrowband,Abbott2019Narrowband,Abbott2022Narrowband}.
Searches around harmonics of the spin
frequency have also been carried out for known pulsars
\citep{Abbott2022TwoHarmonics}.  These are related in spirit, although
do not by themselves cover the quadrupole-weighted frequency
distributions considered here.  While a mismatch of order $10^{-3}$ may be
sufficient for neutron stars within the band of ground-based
interferometers if the poloidal field is sufficiently weak
\citep{mel12}, the key argument here is that signals associated with
magnetic deformations should be spread over a range of frequencies
rather than simply concentrated at ones that are electromagnetically
out-of-phase.

This paper is organised as follows.  We begin in
Section~\ref{sec:gwspect} by exploring the GW spectrum associated with
stars hosting differentially rotating volumes.  The elements behind
magnetic-locking and torque lags are explored in
Section~\ref{sec:diffusion}, with the physical timescales of the
problem estimated in Section~\ref{sec:timescales} for millisecond
pulsars (MSPs; Sec.~\ref{sec:msps}) and magnetars
(Sec.~\ref{sec:magnetars}).  Quantitative applications to magnetic
deformations are considered in Section~\ref{sec:magdeformations} for
constant (Sec.~\ref{sec:cstlags}) or genuinely smeared
(Sec.~\ref{sec:smeared}) spectra, with the impact on source
detectability sketched in Section~\ref{sec:detectability}.  Some
concluding remarks are offered in
Section~\ref{sec:conclusions}, while the appendix discusses microscopic
processes and associated timescales of damping of differential
motions.

\section{Gravitational-wave spectrum} \label{sec:gwspect}

Consider a star in quasi-equilibrium with mass density $\rho$, which
we decompose into two pieces.  First, we have that associated with
rigidly-rotating regions ($\VR$) where the (linear) azimuthal
frequency is $\nR$.  This is the standard setup, where one typically
takes $V_{\rm R}$ to be the whole stellar volume.  In this work,
however, we also allow for a differentially-rotating volume
($V_\Delta$; not necessarily simply-connected), associated with
magnetically-unlocked regions (as in Fig.~\ref{fig:schematic}), where
the star rotates with a position-dependent frequency,
$\nD(\boldsymbol{x})$.  Working in spherical coordinates and an inertial frame,
the total density may be written
$\rho(\boldsymbol{x},t) = \rhoR(r,\theta,\phi - 2\pi\nR t)$
in $\VR$ and
$\rho_\Delta\left(r,\theta,\phi - 2\pi\nD t\right)$
in $\VD$.
  This ansatz assumes that the magnetically induced quadrupolar
  pattern is frozen into local fluid elements, or otherwise maintained
  by magnetic or elastic stresses over the observation time. {This is a
  non-trivial assumption.  Differential rotation winds any poloidal
  components threading regions with different angular velocities,
  generating toroidal field and Maxwell stresses that transmit angular
  momentum on the Alfv{\'e}n or a phase-mixing timescale.  Even if the magnetic
  stress is weak, a spatially-varying pattern speed will tend to mix the
  non-axisymmetric $m=2$ perturbation unless the quadrupole-carrying stress
  can maintain a coherent phase. The formulae introduced here should therefore be
  read primarily as a mapping from an assumed quadrupole-weighted
  pattern-speed distribution to a GW spectrum, not as a proof that such a
  distribution is long lived.}

  A familiar neutron-star example of long-lived differential
    rotation is provided by glitch models, in which the pinned neutron
    superfluid can rotate faster than the electromagnetically coupled
    crust/charged component during
    spindown~\citep{hask15r,hask17}. The lag is then controlled by
    vortex motion and mutual friction, rather than by ordinary viscous
    coupling alone. This example is not identical to the
    magnetic-unlocking scenario considered here, but it illustrates
    that distinct components or regions of a neutron-star interior
    need not corotate on short timescales.

Triaxial equilibria hosting twisted tori were constructed
    by \cite{lm13} showing that, at least in principle,
    non-axisymmetric mass density profiles supported by magnetic
    pressures can exist in neutron-star interiors for arbitrary
    misalignment angles \cite[see also][]{tso22,uryu23}.  There is
    similarly observational evidence that some pulsars
    \cite[such as PSR J0908--4913;][]{kj08} are (near) orthogonal
    rotators, while the interior must consist of mixed poloidal and
    toroidal components to remain stable \citep{akg13}.  There is a
    broader question, however, about whether such non-axisymmetric
    structure can persist over astrophysical timescales when
    differential rotation coexists. {The answer depends on magnetic winding,
    Maxwell stresses, viscosity, and other factors, though simulations by \cite{wei15} suggest there are windows of coexistence (see below).} At least in the context of the Sun, helioseismic
    observations of normal-mode coupling \citep{han19}, Doppler shifts
    of spectral lines \citep{lind97}, and intensity fluctuations
    \citep{duv93} reveal persistent non-axisymmetric substructure. 

In this respect, it is worth highlighting that there exist
  families of exact solutions describing triaxial and
  differentially-rotating stars in the form of the Riemann
  ellipsoids~\citep{chand65}. Such bodies undergo the gravitational
  radiation-reaction driven Chandrasekhar-Friedman-Schutz instability
  which can be damped by normal matter viscosity~\citep{rs20,rs22}.
  In the compressible case, the mass density can be non-axisymmetric
  \citep{lrs93}.  A stability analysis of general-relativistic
  analogues of such compressible ellipsoids was carried out by
  \cite{luo24} for parameter ranges appropriate to neutron stars.  By
  evolving quasi-equilibria for extended timescales ($\gtrsim 30$
  rotational periods), they established the dynamical stability of
  highly-triaxial configurations along sequences of constant rest
  mass.  Notably, they showed that lower-compactness models
  ($C \approx 0.1$) can evolve into states resembling S-type
  ellipsoids that retain substantial non-axisymmetry.  This is
  consistent with the fact that meridional flows are expected in stars
  hosting mixed magnetic fields thanks to circularity theorems
  covering the Einstein-Euler-Maxwell system \citep{bo79,sg23}.
  However, viscosity, radiation reaction, and magnetic fields were not included
  in the simulations by \citet{luo24}, and any of these effects could drive
  the internal flow towards uniform rotation \citep{rs20,rs22,lab26}.
  Moreover, if the degree of differential rotation is too high then
  the star (from the Riemann family or otherwise) will be unstable to
  the bar-mode instability, with the exact threshold depending on the
  equation of state \citep{chand65,shib03}.

  {A related concern is that two non-axisymmetric patterns
    rotating at different rates would continually perturb one
    another. This would indeed be decisive in a no-go sense for two rigid, undeforming
    triaxial bodies sharing a material interface, but that is not
    the picture adopted here.  The relevant quadrupoles exciting GW radiation are moments of a strictly continuous density and stress
    distribution.  Local displacements, induced flows, and elastic/magnetic
    stress adjustments can accompany the relative motion, so
    that the density pattern need not remain fixed up to a common
    interface.  We require a weaker, but still non-trivial condition:
    the non-axisymmetric density or stress variations must retain
    enough phase coherence to generate quadrupole moments with
    distinct pattern speeds over secular timescales.  If the aforementioned effects make the
    phases cancel, the corresponding quadrupole will be
    reduced or redistributed rather than producing a clean shift. 
    An idealised example of this competition is provided by the
    incompressible MHD simulations of \citet{wei15}, who studied
    differentially rotating stars with oblique magnetic fields.  They
    found that differential rotation can symmetrise a weak
    non-axisymmetric field, whereas stronger fields instead brake
    the differential rotation before the non-axisymmetric component is
    erased.  Their dimensionless criteria separating these cases, involving a hierarchy between the fluid
    Reynolds number, effective Alfv{\'e}n
    velocity, and magnetic Reynolds number, are not directly applicable to a stratified, superfluid,
    superconducting star.  Their results, however, illustrate
    that domains of coexistence are not impossible but need to be determined from
    self-consistent resistive, viscous GRMHD simulations.}
  
  Overall, the astrophysical fate of realistic, triaxial stars hosting
  non-rigid flows is a largely open problem.  We do not attempt any
  MHD modelling to address such a problem in this paper: we aim,
  instead, to investigate how such an interplay may impact on
  continuous GW emissions and (eventual) observations.

We start by defining the symmetric trace-free tensor associated with
the position vector, $\hatx = x_i x_j -r^2 \delta_{ij}/3$. Through the
quadrupole-moment tensor,
\begin{equation}
\label{eq:Qij}
Q_{ij}(t) = \int_{V_{\rm R} \cup V_{\Delta}} d^3 \boldsymbol{x} \rho(\boldsymbol{x},t) \hatx ,
\end{equation}
we have the standard, leading-order expression for the GW strain in the transverse-traceless gauge at a distance $d$, viz.
\begin{equation}
\label{eq:hij_t}
h_{ij}(t) = \frac{2G}{c^4 d} \ddot{Q}_{ij}(t),
\end{equation}
where $G$ is Newton's constant, $c$ is the speed of light, and the dot denotes a time derivative \citep{thorne80}.
By expanding the azimuthal dependence of $\rho$ into spatial Fourier modes and keeping only the leading-order terms, it is straightforward to show that expression \eqref{eq:hij_t} becomes
\begin{equation}
\begin{aligned}
\label{eq:hij_explicit}
h_{ij}(t) \approx& -\frac{32 \pi^2 G}{c^4 d} 
\bigg[
\nR^2 \int_{\VR} d^{3}\boldsymbol{x} \hatx \rhoR(\boldsymbol{x}) e^{-4 \pi i \nR t}  \\
&+ \int_{\VD} d^{3}\boldsymbol{x} \nD^2(\boldsymbol{x}) \hatx \rhoD(\boldsymbol{x}) e^{-4\pi i \nD t}
\bigg] .
\end{aligned}
\end{equation}
Notably, since $\nR$ is constant we can pull it outside of the first integral while $\nD$ remains as a weight in the second.

We can Fourier transform expression \eqref{eq:hij_explicit} to understand how a differentially-rotating volume changes the GW appearance of a source. 
Considering an observation time of $2T$, we use the cyclic-frequency convention
\bea
\tilde{h}_{ij}(f)=\int_{-T}^{T} dt\,h_{ij}(t)e^{2\pi i f t},
\eea
where $f$ is measured in Hz. With $\text{sinc}(x)=\sin(\pi x)/(\pi x)$, one has
\bea
\int_{-T}^{T} dt\, e^{2\pi i(f-2\nu)t}=2T\,\text{sinc}\left[2T(f-2\nu)\right],
\eea
for any time-independent rotation frequency $\nu$. This yields
\begin{equation}
\begin{aligned}
\label{eq:hij_omega}
\tilde{h}_{ij}(f) =& \tilde{h}_{i j}^{\mathrm{pol}}(f)+\tilde{h}_{i
    j}^{\mathrm{tor}}(f) \\
\approx&-\frac{64 \pi^2 G T}{c^4 d} 
\bigg\{
\nR^2 \int_{\VR} d^{3}\boldsymbol{x} \hatx \rhoR\,  \text{sinc}\big[{2 T\left(f - 2 \nR\right)}\big]  \\
&+ \int_{\VD} d^{3}\boldsymbol{x} \nD^2\hatx \rhoD \text{sinc}\big[{2 T\left(f - 2 \nD\right)}\big]
\bigg\} ,
\end{aligned}
\end{equation}
where $\tilde{h}_{i j}^{\mathrm{pol/tor}}$ are the respective poloidal and
toroidal components of the tensor (see Sec.~\ref{sec:magdeformations}). As $T \rightarrow \infty$,
  we recover the result corresponding to 
sharp lines at $f=2 \nu_R$ and $f=2 \nu_\Delta$ if both are constant.

{The toroidal radiation result can be generalised by separating the
rotation of the fluid from the pattern speed of the non-axisymmetric
quadrupole.  What determines the GW spectrum is not the geometric volume
distribution of fluid angular velocities, but the quadrupole-weighted
distribution of pattern speeds.  In the simplest frozen-in limit the local
pattern speed equals the local fluid rotation rate,
$\nu_Q(\boldsymbol{x})=\nu_\Delta(\boldsymbol{x})$.  More generally, magnetic
or elastic stresses may support a quadrupolar pattern whose amplitude and
phase are not equivalent to a material interface.  We therefore define
}{
\begin{equation}
W_{i j}(\nu_Q)=\int_{V_{\Delta}} d^3 \boldsymbol{x}\,
\rho_{\Delta}(\boldsymbol{x})\hatx\,\delta\left[\nu_Q-\nu_Q(\boldsymbol{x})\right],
\end{equation}
} {where $\nu_Q$ denotes the pattern speed of the quadrupolar part of
the density or stress distribution.  In the frozen-in approximation used for
the illustrative examples (Sec.~\ref{sec:cstlags}), $\nu_Q=\nu_\Delta$; in a more general
configuration $\nu_Q$ should be understood as the pattern speed entering the
phase of the $m=2$ quadrupole (Sec.~\ref{sec:smeared}).}

{This formulation also clarifies the geometry.  The decomposition into
$V_R$ and $V_\Delta$ should not be read as a cut of the star into two rigid,
non-axisymmetric bodies separated by a material interface.
The boundary of $V_\Delta$ is instead a magnetic or topological boundary, for
example that of a closed-field toroidal zone or a magnetic void, and it may
be axisymmetric or smoothly embedded in the surrounding fluid.  The
non-axisymmetric part of the density perturbation can be localised inside
this region and fall off smoothly towards the boundary.  A schematic two-pattern
representation is
\bea
\delta\rho_{22}(\boldsymbol{x},t)
 =
 A_R(\boldsymbol{x})e^{-4\pi i\nu_R t}
 +
 A_\Delta(\boldsymbol{x})e^{-4\pi i\nu_\Delta t},
\eea
where $A_\Delta(\boldsymbol{x})$ is localised within the magnetically
unlocked region and need not imply a non-axisymmetric interface.
More generally, differential rotation and phase mixing replace the two
idealised pattern speeds by a quadrupole-weighted distribution,
\begin{equation} \label{eq:weightedquad}
Q_{22}(t)
 =
 \int d\nu_Q\,W_{22}(\nu_Q)e^{-4\pi i\nu_Q t}.
\end{equation}
The distribution \eqref{eq:weightedquad}, rather than the rotation of some material sub-body, is
the quantity that determines the continuous-wave spectrum.}
{If viscous or MHD stresses cannot maintain a coherent
non-axisymmetric pattern over the observing time, the corresponding
quadrupole weight $W_{22}$ is reduced or spread over a wider frequency
interval.  In that case a two-frequency model is not applicable; the
signal is instead phase-mixed, smeared, or suppressed.  Constant-lag
examples should therefore be regarded as limiting illustrations, while
the physically relevant description is the quadrupole-weighted pattern-speed
distribution.}

The signal from the differentially rotating region becomes
\bea
\label{eq:hij_general}
\tilde{h}_{i j}^{\Delta}(f) =-\frac{64 \pi^2 G T}{c^4 d} \int d \nu_Q\, \nu_Q^2 W_{i
  j}(\nu_Q) \operatorname{sinc}[2 T(f-2 \nu_Q)] .
\eea
To illustrate the effect of $W_{i j}(\nu_Q)$, suppose most of $V_{\Delta}$ has one pattern speed,
but its density perturbation is nearly axisymmetric there. That region
contributes little to $W_{i j}$, hence little to the GW
signal. Conversely, a small region with a large magnetic distortion
can dominate the spectrum.
In the rigid frozen-in limit, $\nu_Q(\boldsymbol{x})=\nu_\Delta(\boldsymbol{x})=\nu_t$, the quadrupole-weighted
distribution reduces to $ W_{ij}(\nu_Q)=Q^\Delta_{ij}\delta(\nu_Q-\nu_t),$
where $  Q^\Delta_{ij}    =    \int_{V_\Delta} d^3\boldsymbol{x}\,\rho_\Delta(\boldsymbol{x})\hatx$
is the quadrupole moment associated with the unlocked region. Therefore
\begin{equation} \label{eq:hijdeltac}
\tilde{h}_{i j}^{\Delta}(f) = -\frac{64 \pi^2 G T}{c^4 d} \nu_t^2 Q_{i
  j}^{\Delta} \operatorname{sinc}
\left[2T\left(f-2 \nu_t\right)\right] .
\end{equation}
To estimate the extent of smearing, we begin by reviewing the physics of torque lag. 

\section{Diffusion of angular momentum} \label{sec:diffusion}

In most neutron stars, the braking torque associated with the
large-scale dipole dominates the spindown evolution.  Torques applied
to the stellar surface induce boundary (Ekman) layers where the solid
crust meets the liquid core, causing a net fluid transport in the
direction of rotation.  By mass continuity, the horizontal divergence
of the velocity in the Ekman layer drives a vertical pumping flux
which forces a secondary circulation into the core (`Ekman pumping').
Crucially though, the strong stratification gradients in the core
require that any radial flow must overcome buoyant resistance
\citep{mel12}.  Thus, while the primary driver for large-scale
circulation might be torque-induced Ekman pumping, the path of
circulation, including the strength and location of upwelling, is
dictated by the underlying buoyancy and microphysical viscosity.  

On the other hand, the magnetic field is thought to be the fastest
equilibrator \citep{easson79,glamp15}: the field acts like a stiff spring such
that regions linked by poloidal lines relax on the
Alfv{\'e}n timescale, $\tau_{\rm A} \approx L/v_{\rm A}$, where
$v_{\rm A} = B / \sqrt{4 \pi \rho}$ is the Alfv\'en speed, which can
be estimated as
\begin{equation}
 \tau_{\rm A} \sim 35 \left(\frac{L}{10^6
      \mathrm{~cm}}\right)\left(\frac{\rho}{10^{14} \mathrm{~g}
      \mathrm{~cm}^{-3}}\right)^{1 / 2}\left(\frac{10^{12}
      \mathrm{~G}}{B}  \right) \mathrm{~s}.
\end{equation}
Regions of the star that are magnetically-locked in this sense will
essentially rotate uniformly.  However, if there are core regions
that are magnetically decoupled other possible relaxation channels
need to be considered. Candidates of such magnetically decoupled
regions are toroidal zones enclosed by neutral curves or magnetic voids. 
The absence of a magnetic field can be due to local
superconducting Meissner expulsion; in fact, the twisted torus
configurations of superconducting neutron stars render the structure
and topology of superconducting regions quite
complicated~\citep{Das2025,Das2026}.  Several mechanisms contribute to
the rotational relaxation; examples of such are diffusion by
thermobuoyant ($\tau_{\kappa}$) or viscous ($\tau_{\eta}$) processes
\citep{mel12}. Furthermore, superfluid mutual friction can contribute
to recoupling between the core and the crust; however it may be
ineffective and may lead to long-term relaxation on secular
timescales~\citep{Sedrakian1998}. MHD instabilities,
ambipolar drift, or Hall-mediated rearrangement could be a
factor. Finally, the compositional diffusion (on timescale
$\tau_{\beta}$) can play a r{\^o}le, as discussed in the next section.

A useful way of parameterising the lag is through a relaxation
solution, 
\begin{equation} \label{eq:deltanuformula} 
\Delta \nu \sim -\tau_{\mathrm{eff}} \dot{\nu}_{\rm R}, 
\end{equation} 
for steady spindown,
provided that the relaxation time is shorter than the spindown age.
More generally, $\tau_{\rm eff}$ should be understood as the time
associated with complete torque channels acting on the
magnetically-unlocked region.  We therefore write schematically
\bea\label{eq:tau_eff} \tau_{\mathrm{eff}}^{-1}\simeq
\tau_B^{-1}+\tau_{\eta,\perp}^{-1}+\tau_{\kappa}^{-1}
+ \left[\tau_{\rm ES}^{(\beta)}\right]^{-1}+\tau_{\rm mf}^{-1}
+\cdots .  \eea Here
$\tau_B$ denotes magnetic coupling across the boundary of the unlocked
region, $\tau_{\eta,\perp}$ is the viscous diffusion time for
angular-momentum transport across magnetic surfaces, $\tau_\kappa$ is a
thermal-diffusion, or thermal Eddington--Sweet, channel,
$\tau_{\rm ES}^{(\beta)}$ is the composition-limited Eddington--Sweet
circulation time, and $\tau_{\rm mf}$ is a possible superfluid
mutual-friction coupling time. In a region not connected to the crust
by poloidal field lines, such as a closed-field toroidal zone or a
magnetic void, the usual Alfv\'enic magnetic torque channel is
strongly suppressed. We therefore treat $\tau_B$ as effectively
infinite, or at least much longer than the ordinary Alfv\'en crossing
time.  The beta-reaction time should not be added separately to
Eq.~\eqref{eq:tau_eff}; it is a necessary sub-step of the
composition-limited circulation channel and is already included in
$\tau_{\rm ES}^{(\beta)}$.

In the above, $\dot{\nu}_{\rm R}$ is the spindown rate determined by
the braking torque for a star with polar-dipole strength $B_{p}$~\citep{spit06},
\begin{equation} \label{eq:nuformula}
\dot{\nu}_{\rm R}
\approx -\frac{\pi^2 B_{p}^2 R^6 \nR^3}{I_0 c^3}(1 + \sin^2\alpha),
\end{equation}
where $\alpha$ is the angle between the magnetic and rotation axes
(Fig.~\ref{fig:schematic}) and we ignore relativistic corrections
\cite[see, e.g.,][]{stef25}.
We take $\alpha=\pi/2$ in the representative estimates in order to maximise the
$m=2$ contribution at $2\nR$. Other inclinations change the relative
amplitudes and may introduce additional harmonics \citep{lm13}, but they do not
alter the basic spectral-smearing mechanism.  Henceforth, we also fix
$M = 1.4 M_{\odot}$ and $R = 12$~km to avoid
carrying around factors of mass and radius.

Taking $I_{0} \approx 2.2 \times 10^{45} \text{ g}~\text{cm}^2$ for the moment of inertia, we adopt the numerical estimate
\begin{equation} \label{eq:omegadot}
\dot{\nu}_{\rm R} \approx -9.6 \times 10^{-16}
\left( \frac {B_{p}}{10^{12} \text{ G}}\right)^{2}
\left( \frac{\nR}{\text{1 Hz}} \right)^3 \text{ Hz s}^{-1},
\end{equation}
though none of our conclusions change for a slightly different normalisation (a lower $I_{0}$ would only increase the lag in Eq.~\ref{eq:deltanuformula}).
Associated with
Eq.~\eqref{eq:omegadot} is the characteristic spindown time,
\begin{equation} \label{eq:spindown}
    \tau_{\rm sd} = \frac{\nu_R}{2|\dot\nu_R|}
    \approx 1.7 \times 10^7
    \left( \frac {B_{p}}{10^{12} \text{ G}}\right)^{-2}
    \left( \frac{\nR}{\text{1 Hz}} \right)^{-2}
    \text{ yr}.
\end{equation}
For the fiducial MSP values used below,
$B_p=10^9\,{\rm G}$ and $\nu_R=500\,{\rm Hz}$, this gives
$
    \tau_{\rm sd}\simeq 7\times10^7\,{\rm yr}.
$
For a magnetar-like source with $B_p=10^{14}\,{\rm G}$ and
$\nu_R=0.1\,{\rm Hz}$, one instead obtains
$
    \tau_{\rm sd}\simeq 1.7\times10^5\,{\rm yr}.
$
Thus the electromagnetic spindown time is still secular in both cases,
but it is substantially shorter for magnetars.

\section{Physical interpretation of the coupling timescales} \label{sec:timescales}

The persistence of differential rotation in magnetically-unlocked regions is controlled by the torque channels in Eq.~\eqref{eq:tau_eff}.  The short Ekman time is not, by itself, the global corotation time in a strongly stratified neutron-star core.  Rather, as emphasised by \cite{mel12}, Ekman pumping rapidly establishes a stratification-confined sheared state, while the subsequent removal of the shear is governed by slower secular processes.  For the present problem there are two additional ingredients.  First, the relevant viscous coefficient for a toroidal magnetic region is not necessarily the zero-field isotropic shear viscosity, but the transverse component of the magnetised viscosity tensor.  Second, if the neutron component is paired, the neutron quasiparticle contribution can be exponentially suppressed.  We therefore treat the long-lived lag as a parameter-dependent possibility rather than as an unavoidable consequence of compositional stratification.

\subsection{A composition-controlled secular channel} \label{sec:eddsweet}

A meridional circulation that carries fluid elements across a compositionally stratified background is limited by the ability of weak interactions to restore beta equilibrium.  Following the notation of \cite{mel12}, a composition-controlled Eddington--Sweet analogue may be written
\begin{equation}
    \tau_{\rm ES}^{(\beta)} \;\sim\; S\,\tau_\beta,\quad S = \frac{N^2}{\Omega^2},
  \end{equation}
where $N$ is the Brunt--V\"ais\"al\"a frequency and $\Omega =2 \pi \nR$ is the angular spin frequency. The stiffness parameter is given by~\citep{mel12}
\begin{equation}\label{eq:S}
    S \;\simeq\; 2.9\times10^{2}
    \left(\frac{\rho}{\rho_0}\right)^{1/3}
    \left(\frac{\Omega}{10^2\;\mathrm{s}^{-1}}\right)^{-2},
\end{equation}
where $\rho_0  = 2.8\times10^{14}\,{\rm g\,cm^{-3}}$ is the nuclear saturation density.

For modified-Urca reactions, using the results of \cite{Goldreich1992} and \cite{Reisenegger1995},
\begin{equation}\label{eq:taubeta_MU}
    \tau_\beta^{\rm MU}
    \;\sim\;
    5\times10^3
    \left(\frac{\rho}{\rho_0}\right)^{-1/3}
    \left(\frac{T}{10^8\;\mathrm{K}}\right)^{-6}
    \;\mathrm{yr}.
  \end{equation}
Combining this result with Eq.~\eqref{eq:S} gives
\begin{equation}\label{eq:tES_beta_corrected}
    \tau_{\rm ES}^{(\beta)}
    \;\sim\;
    1.5\times10^{6}
    \left(\frac{T}{10^8\;\mathrm{K}}\right)^{-6}
    \left(\frac{\Omega}{10^2\;\mathrm{s}^{-1}}\right)^{-2}
    \;\mathrm{yr}.
  \end{equation}
The combined density dependences in $S$ and $\tau_\beta^{\rm MU}$ cancel at this level of approximation.  Equation~\eqref{eq:tES_beta_corrected} should be interpreted as the timescale of one complete secular channel: beta-limited, composition-controlled circulation.  It does not mean that beta equilibration is a universal bottleneck for every possible angular-momentum-transport mechanism.  Magnetic coupling, superfluid mutual friction, or direct viscosity can dominate if the corresponding torque channel acts efficiently on the quadrupole-carrying component.

The apparently fast viscous channel requires particular care in the
present geometry.  The differential flow is mainly azimuthal, whereas
the angular-momentum flux needed to erase the lag is across the
magnetic surfaces enclosing the toroidal zone.  The relevant
coefficient is therefore the transverse viscosity $\eta_\perp$.
Magnetised-core transport calculations show that the shear-viscosity
tensor becomes strongly anisotropic when the electron magnetisation
parameter is large \citep{ShterninOfengeim2022}.  In addition, the
neutron quasiparticle viscosity is suppressed by ${}^3P_2$--${}^3F_2$
pairing; microscopic calculations summarised in Fig.~22 of
\cite{SedrakianClark2019} give representative gaps $\Delta_n$ of
several tenths of an MeV around the relevant neutron Fermi wave-number
$k_{Fn}\simeq1.7~{\rm fm}^{-1}$, although the uncertainty remains
substantial.  Appendix~\ref{sec:appendix} shows that, for
magnetar-like parameters with $B_{\rm tor}\sim10^{14}$~G,
$T\simeq6\times10^8$~K, and a gap near the upper part of this range,
$\Delta_n\simeq0.6$~MeV, one finds
\begin{equation}
    \tau_{\eta,\perp}(L_\Delta\simeq0.1R)
    \sim {\rm few}\times10^6~{\rm yr},
\end{equation}
whereas Eq.~\eqref{eq:tES_beta_corrected} gives
\begin{equation}
    \tau_{\rm ES}^{(\beta)}\sim8\times10^5~{\rm yr}.
\end{equation}
Thus direct viscous coupling need not be the fastest secular channel
in the strongly magnetised, strongly paired regime.  For smaller gaps,
for example $\Delta_n\simeq0.2$~MeV, the residual neutron viscosity
remains large and viscous relaxation is much faster.  The existence of
a long-lived lag is therefore dependent on the microphysics of pairing, magnetised transport, and mutual friction.

For steady spindown and $\tau_{\rm eff}$ shorter than the spindown age, Eq.~\eqref{eq:deltanuformula} gives the usual quasi-stationary lag.  If instead $\tau_{\rm eff}$ exceeds the spindown age, the star retains a fossil differential rotation set by its earlier spin history, and $-\tau_{\rm eff}\dot\nu_R$ should be regarded only as a formal scale rather than a steady-state prediction.

\subsection{Millisecond pulsars} \label{sec:msps}

The size of the lag is highly sensitive to the stellar temperature,
magnetic field strength, spin frequency, and to which coupling
mechanism dominates angular-momentum exchange.  Ultraviolet and
soft-X-ray observations of the $\gtrsim$~Gyr old object J0437--4715
indicate thermal emission from the stellar surface, with redshifted
effective temperatures of order $\sim 2 \times 10^5$~K \citep{karg04,GonzalezCaniulef2019}.
Such temperatures in MSPs are generally interpreted
as evidence for internal reheating, for example by rotochemical
heating \citep{FernandezReisenegger2005,GonzalezJimenez2015}
or vortex creep \citep{VanRiper1994,Schaab1999}.
The relationship between the surface and core
temperatures is therefore model dependent \citep{pot01}; in the illustrative estimates below we adopt
$T=3\times10^7$~K as a fiducial cold-core value. 

For a $500$~Hz pulsar, Eq.~\eqref{eq:S} gives $S\simeq0.3$ at
$\rho=\rho_0$.  This is not a strongly stratified regime, and the
beta-limited Eddington--Sweet time should not be expected to set
$\tau_{\rm eff}$ unless faster channels are suppressed.  For comparison
with the figures below, we nevertheless quote the formal lag obtained by
inserting the beta-limited composition-controlled scale
\eqref{eq:tES_beta_corrected} into Eq.~\eqref{eq:deltanuformula}:
\begin{equation} \label{eq:rossest}
\Delta \nu_{\rm ES}^{(\beta)} \approx 7.9 
\left( \frac {B_{p}}{10^{9} \text{ G}}\right)^{2}
\left( \frac{\nR}{\text{500 Hz}} \right)
\left(\frac{T}{3 \times 10^7 \text{ K}} \right)^{-6}
\text{Hz}.
\end{equation}
This value is an upper-bound, or long-coupling, reference point rather than an
average-parameter MSP prediction.  At the fiducial point it
corresponds to $\tau_{\rm ES}^{(\beta)} \simeq 2.1$~Myr,
whereas the transverse electron-viscous time from
Eq.~\eqref{eq:tau_eta_eperp} is much shorter for the same temperature,
\begin{equation}
\tau_{\eta,e\perp}
\simeq 40~{\rm yr},
\end{equation}
for $B_{\rm tor}=10^9$~G and $L_\Delta=R$, giving
$\Delta\nu \sim 0.1$~mHz.  Even for
$B_{\rm tor}=10^{11}$~G, one finds
\begin{equation}
\tau_{\eta,e\perp}
\simeq 4\times10^5
\left(\frac{L_\Delta}{R}\right)^2~{\rm yr},
\end{equation}
so that a localised region with $L_\Delta=0.1R$ gives only
$\Delta\nu\sim10^{-2}$~Hz.  The normal-proton thermal
Eddington--Sweet channel in Appendix~\ref{sec:appendix} is even shorter:
Eq.~\eqref{eq:tkappa_normal} gives $\tau_\kappa^{(N)}\sim$ a few years,
while Eq.~\eqref{eq:tkappa_normal_leptons} gives about $1$~yr in the
lepton-heat-capacity limit.  If this channel acts as a complete angular-momentum-transport channel, the corresponding lag is only
$\Delta\nu\sim10^{-5}$~Hz.

Thus a Hz-scale lag in an MSP requires the relevant viscous, thermal,
magnetic, and mutual-friction coupling channels to be strongly
suppressed.  Equivalently, if the lag were controlled by mutual friction
alone, the required coefficient would be
\begin{equation}
\tilde{\mathcal{B}}
\sim (2\Omega\tau_{\rm eff})^{-1}
\sim 10^{-18},
\end{equation}
far below standard core estimates \citep[e.g.,][]{hask15r,hask17}.  In
addition, the neutron superfluid is not confined by the magnetic topology
in the same way as the charged, field-carrying component; efficient
vortex-mediated coupling would bridge closed- and open-field regions and
reduce the lag.  The MSP examples below using expression \eqref{eq:rossest} should therefore
be understood as phenomenological tests of the spectral mapping, rather
than as robust predictions backed up by microphysics.

\subsection{Magnetars} \label{sec:magnetars}

In the opposite regime corresponding to Galactic magnetars, the
rotation is slow and the internal magnetic field can be strong enough
for both stratification and magnetic anisotropy of transport to
matter.  For a young magnetar with an age of order kyr,
$\nR \approx 0.1$~Hz, and $B_{p} \gtrsim 10^{14}$~G, one expects
a core temperature of order
$T_{\rm core} \approx 6 \times 10^{8}$~K
\citep{Beloborodov16}. If the effective secular coupling time is comparable to the estimate in
Eq.~\eqref{eq:tES_beta_corrected}, the formal steady-state scale would be
\begin{equation} \label{eq:magnetarross}
\Delta \nu_{\rm ss}^{(\beta)} \approx 0.25
\left( \frac {B_{p}}{10^{14} \text{ G}}\right)^{2}
\left( \frac{\nR}{\text{0.1 Hz}} \right)
\left(\frac{T}{6 \times 10^8 \text{ K}} \right)^{-6}
\text{Hz}.
\end{equation}
This value should not be interpreted as the spindown lag accumulated by a
young magnetar of age $t_{\rm age}\sim {\rm kyr}$.  For the fiducial
parameters, Eq.~\eqref{eq:tES_beta_corrected} gives
$\tau_{\rm ES}^{(\beta)}\simeq8\times10^5$~yr, which is longer than
both the age and the electromagnetic spindown time
$\tau_{\rm sd}\simeq1.7\times10^5$~yr.  The quasi-steady approximation
is therefore not reached.  If the unlocked region were initially corotating
with the crust, the age-limited lag would instead be
\begin{equation}
\Delta\nu_{\rm age}
\simeq |\dot{\nu}_{R}|t_{\rm age}
\simeq
3\times10^{-4}
\left(\frac{t_{\rm age}}{10^3\,{\rm yr}}\right)
{\rm Hz}.
\end{equation}
A lag as large as Eq.~\eqref{eq:magnetarross} must therefore be interpreted
as a fossil lag, set by the earlier spin history of the star, and is bounded
by
\begin{equation}
\Delta\nu_{\rm foss}\lesssim \nu_{\rm birth}-\nu_R .
\end{equation}
The examples in Sec.~\ref{sec:magdeformations} use this large value only as
an illustrative upper bound for the spectral mapping.  

In any case, we anticipate a smearing from $2\nR$ to $2 (\nR + \Delta \nu)$ in the GW spectrum for mass quadrupoles, which we explore next.

\section{Magnetic deformations} \label{sec:magdeformations}
Magnetic stresses are expected to deform neutron stars away from
spherical symmetry \citep{c53}.  Even if the deformation is small, the
induced ellipticity, a signed proxy for the mass quadrupole moment,
may be large enough to be detectable by (near-future) GW
interferometers \citep{wette23}. 
In a star hosting a mixed poloidal-toroidal field, the two partitions
induce ellipticities of
\begin{equation} \label{eq:epsilonpol}
    \epsilon_{\rm pol} \sim E_{\rm pol}/E_{G}, \qquad \epsilon_{\rm tor} \sim -E_{\rm tor}/E_{G},
\end{equation}
where $E_{\rm pol,tor}$ denote the magnetic energies stored in the
respective components and $E_{G}$ is the gravitational binding energy
\citep{cut02}.  Note, importantly, that $\epsilon_{\rm pol}$ and
$\epsilon_{\rm tor}$ have opposite signs and that the numerical
coefficients implied by expression \eqref{eq:epsilonpol} match to
within a factor of a $\sim$~few to calculations using realistic
equations of state \citep{fr12,gom19}, at least for large-scale fields
\citep{mast15b}.  The exact values are unimportant for our purposes,
as our main goal is to illustrate spectral smearing.  Note that an
axisymmetric deformation emits continuous GWs only if its symmetry
axis is not aligned with the rotation axis (see
Fig.~\ref{fig:schematic}), or if the star is triaxial. Otherwise a
purely axisymmetric deformation rotating about its own symmetry axis
gives no $m=2$ emission.  Moreover, the relative amplitudes
  of the two contributions are controlled by the integrated quadrupole
  moments, not by the field amplitudes alone.  In the magnetic
  estimate this dependence enters through
\begin{equation} 
E_a=\int_{V_a}\frac{B_a^2}{8\pi}\,dV,
\qquad a={\rm pol,tor},
\end{equation} 
together with geometry-dependent coefficients. Thus the volume,
location, and concentration of the magnetic stresses are all
implicitly contained in $\epsilon_{\rm pol}$ and $\epsilon_{\rm
  tor}$. In the examples below we therefore vary the ratio
$E_{\rm tor}/E_{\rm pol}$, rather than assuming that a field-strength
ratio alone fixes the quadrupole ratio.

If the toroidal field is confined to neutral curves within the star
\citep{bra06,lm13}, we anticipate {core} regions associated with
$\epsilon_{\rm tor}$ to rotate at a rate of
$ \nD = \nR + \langle \Delta \nu \rangle$ on average, while poloidal
regions rotate at the surface rate, $\nR$.  To get some first
(extreme) estimates, suppose that $\nD$ is constant and
slowly-varying within the toroidal volume so that we can apply
expression \eqref{eq:hijdeltac}.  In terms of the ellipticities, we
get an explicit formula for the strain in the form
\begin{equation} \label{eq:exampleh0}
\begin{aligned}
\tilde{h}_{0}(f) \propto& \big|\epsilon_{\rm tor} \left(\nR+\Delta \nu\right)^2 \text{sinc}[2T(f - 2\nR - 2\Delta \nu)] \\
    &+\epsilon_{\rm pol} \nR^2 \,\text{sinc}[2T(f - 2\nR)] \big|.
    \end{aligned}
  \end{equation}
  In the $\Delta \nu \to 0$ limit, the sum collapses and the two
  deformations compete: the poloidal field works to make the star
  oblate and the toroidal one to make it prolate, with
  $|\epsilon_{\rm pol} + \epsilon_{\rm tor}| < |\epsilon_{\rm pol}| +
  |\epsilon_{\rm tor}|$.  If the toroidal field is unlocked though,
  the pieces enter the strain \eqref{eq:exampleh0} in a decoupled
  way. Note that we used scalar addition in equation
  \eqref{eq:exampleh0} assuming the same $m=2$ projection,
  orientation, principal axis, and so on.  More generally, the two
  quadrupoles are tensors and
  $\tilde{h}_{i j}(f)=\tilde{h}_{i j}^{\mathrm{pol}}(f)+\tilde{h}_{i
    j}^{\mathrm{tor}}(f),$ whereby the scalar expression arises as a
  simplified axis-aligned model.
{In all spectra shown in this paper, the absolute vertical scale is arbitrary
because the strain scales inversely with distance. Within each panel, however, the relative
poloidal and toroidal amplitudes are fixed using the magnetic-ellipticity
estimate of \citet{cut02}, with $\epsilon_{\rm pol}$ and
$\epsilon_{\rm tor}$ evaluated from the quoted
$\langle B_{\rm pol}\rangle$  and $\langle B_{\rm tor}\rangle$
values for the illustrative geometry.  The detailed amplitudes would
change with the magnetic geometry, field localisation, and lag, but the plotted
ratios provide a concrete example.}

\subsection{Constant lags} \label{sec:cstlags}

\begin{figure}
\includegraphics[width=0.465\textwidth]{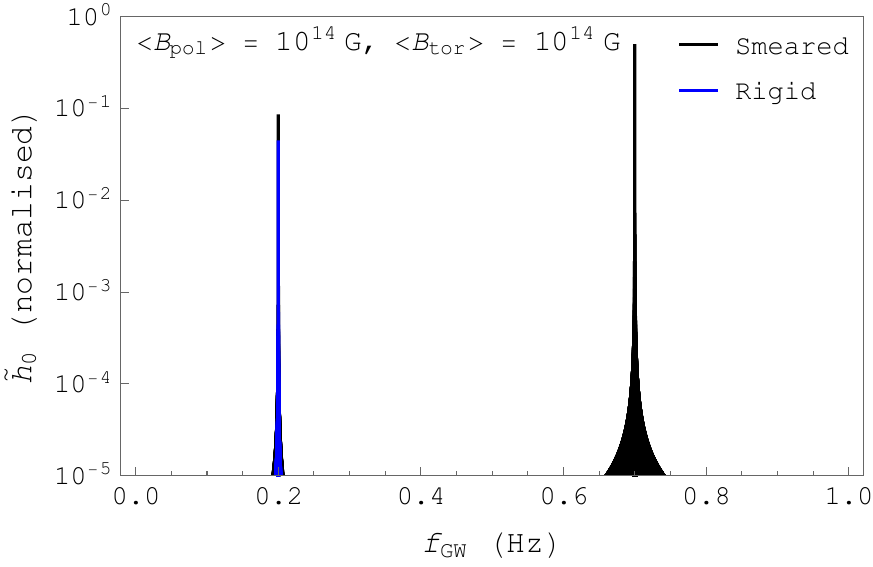}
\includegraphics[width=0.465\textwidth]{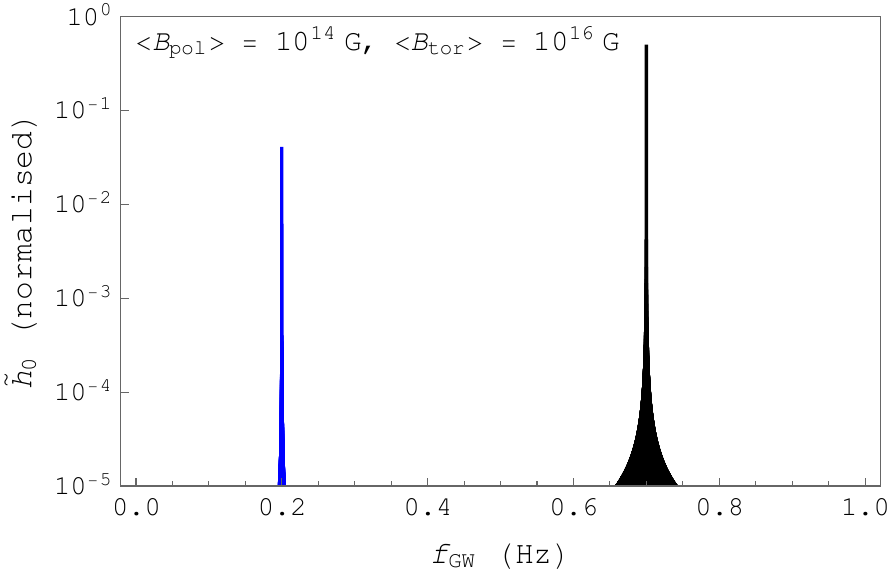}
\caption{GW spectra in cases of constant torque-lag in toroidal
  regions for magnetar-like parameters, using the deliberately large
  fossil-lag example  discussed in the text. 
Black curves show the expectations with torque lag, while blue ones assume instead the entire star rotates at $\nR = 0.1$~Hz.
The scale for $\tilde{h}_{0}$ has been normalised to different values
across the panels to highlight relative magnitudes rather than to
provide direct estimates for the strain itself.
\label{fig:magnetars}}
\end{figure}

We first consider, for illustration, spectra produced by a constant lag.
This is an idealised limiting case: in a realistic fluid, shear instabilities
and secular coupling processes are expected to smooth sharp rotational
jumps at the boundaries. Nevertheless, the constant-lag model is useful
because it gives the maximum peak separation in Fourier space.
In what follows, $\Delta\nu$ denotes the lag, while
$\nu_\Delta=\nu_R+\Delta\nu$ denotes the total rotation rate, or pattern
speed, of the unlocked quadrupole-carrying region.

Taking $2T=1$ month, two examples of a normalised $\tilde h_0(f)$
from Eq.~\eqref{eq:exampleh0} are shown in Fig.~\ref{fig:magnetars}
for magnetar-like parameters, $B_p=10^{14}$~G and
$\nu_R=0.1$~Hz. The upper panel shows the case
$E_{\rm tor}=E_{\rm pol}$, while the lower panel shows a toroidally
dominated case. In these panels we deliberately use the large fossil-lag
example $\Delta\nu_{\rm foss}=0.25$~Hz, comparable to the formal value
in Eq.~\eqref{eq:magnetarross}, in order to display the maximum spectral
separation.  This should not be interpreted as the age-limited lag of a kyr-old
magnetar that was initially corotating.  For such a case the expected
spindown-generated lag would be orders of magnitude smaller.

With this illustrative choice, the signal is no longer centred only at
$2\nu_R$. A second contribution appears at
$2\nu_\Delta=2(\nu_R+\Delta\nu)$.  For
$\Delta\nu_{\rm foss}=0.25$~Hz one has
$\nu_\Delta\simeq0.35$~Hz and $\nu_\Delta/\nu_R\simeq3.5$.
Because the strain contribution scales as the square of the pattern
frequency, the shifted toroidal component is enhanced relative to a
component with the same quadrupole moment rotating at $\nu_R$. In the
toroidally dominated case, the higher-frequency peak therefore dominates
the spectrum.

We conclude that if the toroidal region rotates faster than
magnetically locked ones and most of the magnetic energy is contained
there, then most of the power can lie outside the conventional bin.
This is a consequence of the spectral mapping of an assumed
quadrupole-weighted pattern-speed distribution; we do not claim that
young magnetars are generically bound to develop such large spindown lags.

Although Galactic magnetars are usually far below the sensitive band of
ground-based interferometers, moving a $0.1$~Hz magnetar into the
$\sim10$~Hz GW band would require
\begin{equation}
\Delta\nu  \sim 5~{\rm Hz},
\end{equation}
that is, an unlocked quadrupole pattern rotating more than ten times
faster than the observed surface.  Such a configuration would require a
large fossil spin reservoir and is not claimed here as a realistic
prediction.  The associated excess rotational energy is approximately
\begin{equation}
\Delta E_{\rm rot}
\simeq
2\pi^2 I_\Delta
\left[(\nu_R+\Delta\nu)^2-\nu_R^2\right].
\end{equation}
For the illustrative value $\Delta\nu=0.25$~Hz and $\nu_R=0.1$~Hz this is
\begin{equation}
\Delta E_{\rm rot}
\simeq
2.2\times10^{45}
\left(\frac{I_\Delta}{10^{45}\,{\rm g\,cm^2}}\right)
{\rm erg},
\end{equation}
whereas for $\Delta\nu=10$~Hz one obtains
\begin{equation}
\Delta E_{\rm rot}
\simeq
2.0\times10^{48}
\left(\frac{I_\Delta}{10^{45}\,{\rm g\,cm^2}}\right)
{\rm erg}.
\end{equation}
It is unlikely therefore that such lags could push magnetars into the ground-based band.

\begin{figure}
\includegraphics[width=0.465\textwidth]{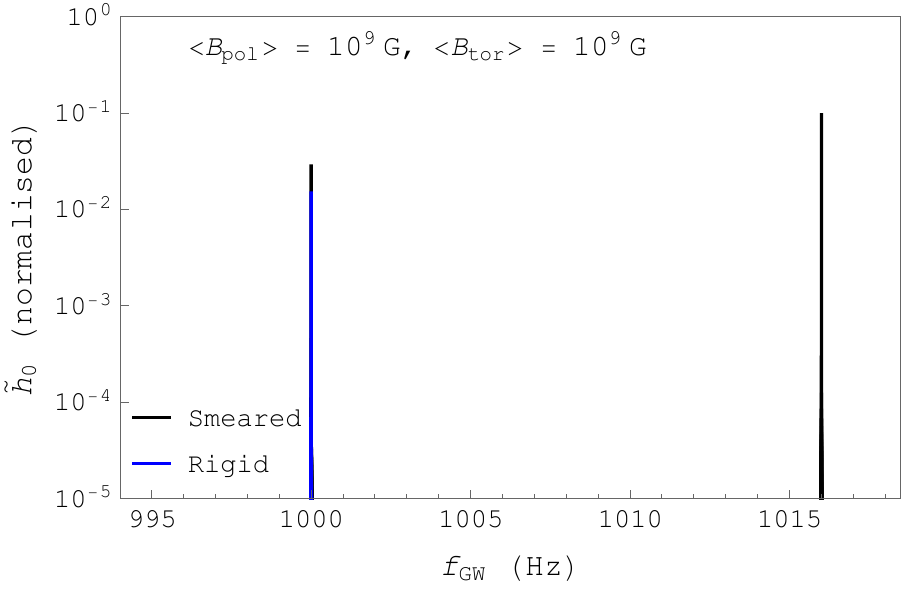}
\includegraphics[width=0.465\textwidth]{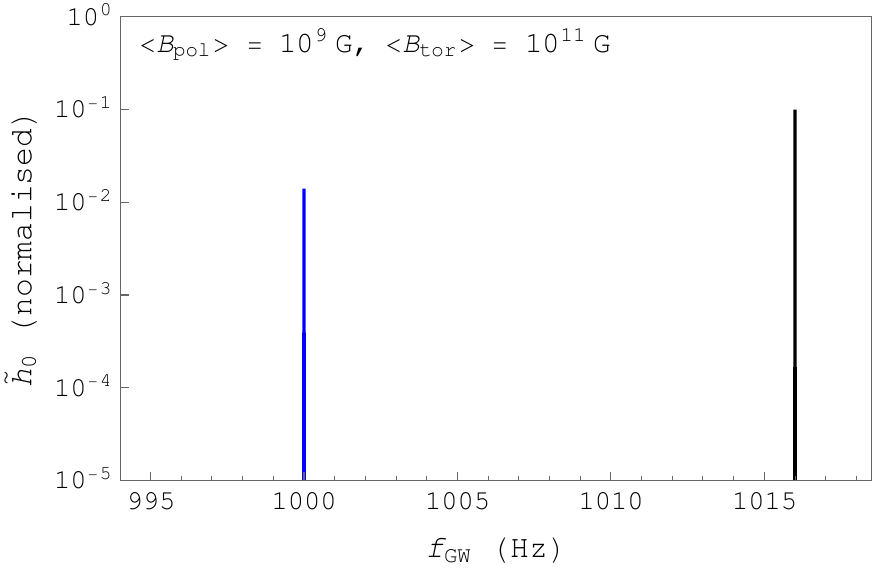}
\caption{Similar to Fig.~\ref{fig:magnetars} but for $\nR = 500$~Hz
  and weaker fields, using the deliberately large beta-limited lag
realisation of Eq.~\eqref{eq:rossest}. 
  \label{fig:msps}}
\end{figure}

Figure~\ref{fig:msps} shows the corresponding constant-lag spectra for
MSP parameters, $B_{p}=10^{9}$~G and $\nR=500$~Hz,
using the deliberately large beta-limited lag scale in
Eq.~\eqref{eq:rossest}.  This choice illustrates how a
quadrupole-weighted pattern-speed offset would appear in a
continuous-wave spectrum, but it should not be interpreted as a generic
MSP prediction.  As discussed in Sec.~\ref{sec:msps} and
Appendix~\ref{sec:appendix}, standard viscous, thermal, or
mutual-friction coupling estimates may well reduce the lag
below this value.

In this large-lag example the shifted toroidal
contribution lies outside the usual narrowband window around $2\nu_R$ \cite[e.g.,][]{abb08}.
For standard MSP coupling estimates, however, the lag may be much
smaller and may remain inside the searched band.  The robust conclusion
is therefore not that recycled pulsars generically evade narrowband
searches, but that a search centred only at $2\nu_R$ constrains only the
quadrupole weight lying within the assumed frequency window.  Because
the strain contribution scales as $\nu^2$, any shifted component is also
enhanced or reduced according to its own pattern frequency.

Even without smearing (explored below) the results obtained here have a
number of implications:
\begin{itemize}
    \item[(i)] {oblate and prolate distortions cannot `conspire' to produce an overall small quadrupole moment;}
        \item[(ii)] {GW detections and/or upper limits can constrain the {individual} toroidal and poloidal components, unlike in the everywhere rigidly-rotating case;}
        \item[(iii)]{single-template or narrowband searches centred around $2 \nR$ constrain only the quadrupole weight that lies inside the assumed frequency window, and may miss power if a non-negligible part of the pattern-speed distribution is shifted or smeared, and}
        \item[(iv)]{uncertainties on inferred nuclear or magnetic parameters increase in the event of a detection because the strain \eqref{eq:exampleh0} depends on the pattern-speed distribution, which is sensitive to temperature, magnetic topology, pairing, and the active coupling channels.}
\end{itemize}

\subsection{Smeared spectra} \label{sec:smeared}

\begin{figure}
\includegraphics[width=0.47\textwidth]{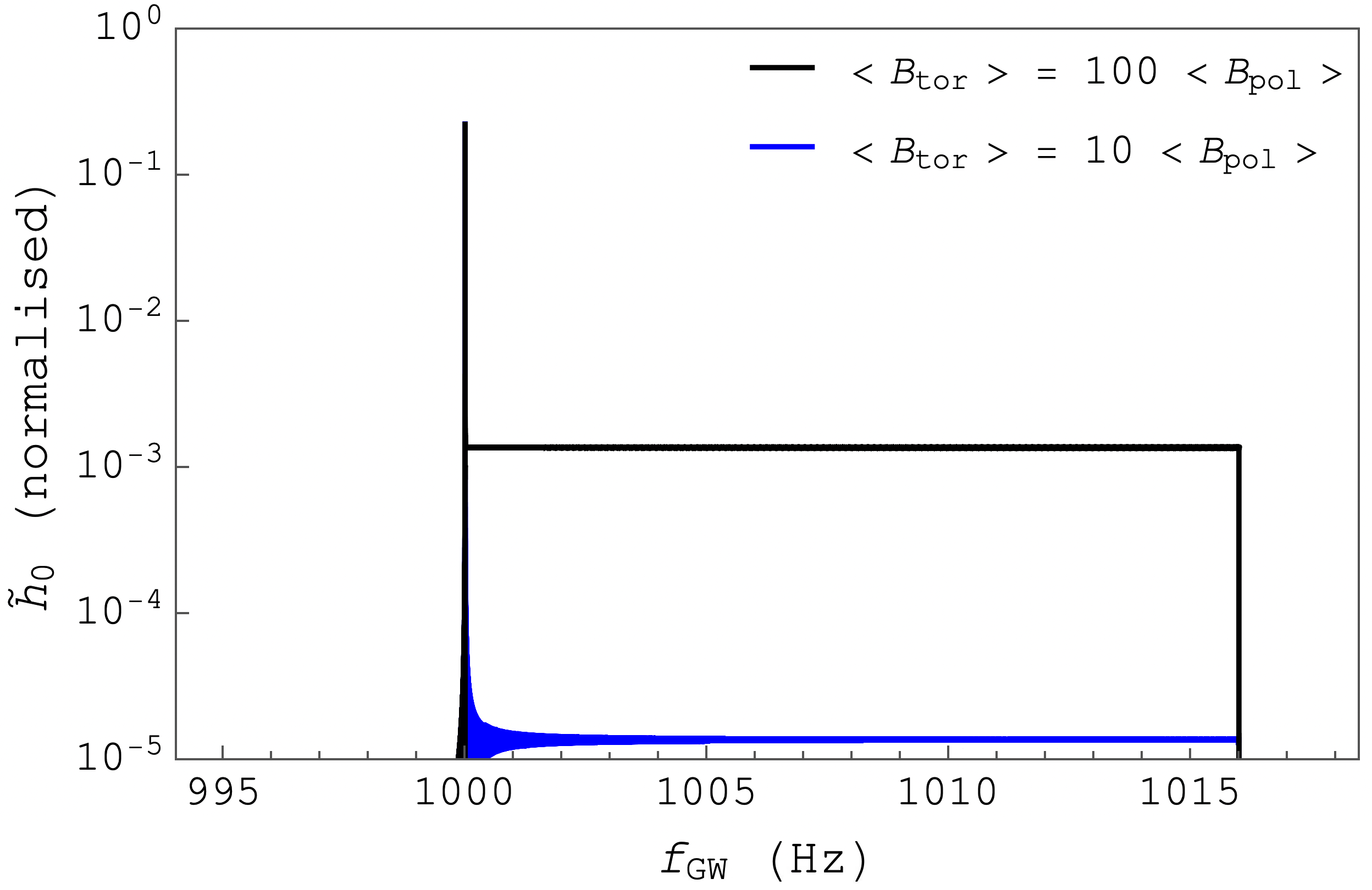}
\caption{Smeared spectra for the millisecond-pulsar-like parameters of
  Fig.~\ref{fig:msps}, using the same deliberately large-lag example 
  and an illustrative uniform quadrupole-weighted pattern-speed
  distribution across the toroidal-frequency interval. The observation
  time is $2T=1$ month.}
\label{fig:magsmears}
\end{figure}

We next consider a more realistic scenario in which $\nu_\Delta$
varies throughout the magnetically-unlocked region.  In general, a
constant $\nD$ is likely to invite shear instabilities at the toroidal
boundary and non-axisymmetries are unlikely to persist over even
dynamical timescales.  Cases with a spatially varying $\nD$ are more
plausible and indeed exact, albeit simplified, configurations can be
constructed \citep{lai93,wei15,luo24}.  Predicting a realistic profile
requires solving the Navier-Stokes equations coupled to that
describing thermal relaxation; this is a formidable task that we do not
attempt here.  We can still get an idea for the smearing however by
noting that, in the axisymmetric limit, Ferraro's theorem requires
that the angular velocity is just a function of the poloidal flux,
$\psi$ \cite[see, e.g.,][]{sg23}.  Integrals over the $\text{sinc}$
function thus effectively become one-dimensional after transforming
into magnetic coordinates oriented along contours of constant $\psi$.

For illustration we adopt a uniform quadrupole-weighted distribution
$W_{ij}(\nu_Q)$ over the interval
$\nu_R\le \nu_Q\le \nu_R+\Delta\nu$ in
Figure~\ref{fig:magsmears}.  This should not be confused with assuming
only a linear spatial profile for $\nu_\Delta(\boldsymbol{x})$.  In a
realistic toroidal geometry the volume element, the density perturbation,
and the tensor weight $\hat X_{ij}$ all enter the mapping from position
to pattern speed.  The function $W_{ij}(\nu_Q)$ would therefore have to
be computed from a self-consistent MHD model.

For a uniform quadrupole-weighted distribution over a rotational-frequency
interval $\Delta\nu$, the GW bandwidth is just $\Delta f_{\rm GW}\approx 2\Delta\nu$ .
If the total quadrupole weight in the smeared component is fixed, the
coherent Fourier amplitude in a single bin is reduced
relative to an unsmeared line by approximately
\begin{equation} \label{eq:perbin_reduction}
    \frac{|\tilde h_{\rm bin}|}{|\tilde h_{\rm line}|}
    \sim
    \frac{Q_\Delta}{Q_{\rm pol}}
    \frac{1}{\Delta f_{\rm GW}T}
    \sim
    \frac{Q_\Delta}{Q_{\rm pol}}
    \frac{1}{2\Delta\nu T},
\end{equation}
up to factors of order unity that depend on the precise definition of the kernel and binning convention.  For the deliberately large
MSP limiting example, $\Delta\nu=7.9$ Hz and $2T=1$ month, even
$Q_\Delta/Q_{\rm pol}\sim 10^2$ gives a local amplitude ratio of only
order $\sim 10^{-5}$.  This is why the smeared spectra shown
below use different normalisation from 
Eq.~\eqref{eq:perbin_reduction}, which gives the physical local per-bin scaling.

With this caveat, the qualitative interpretation remains useful.  If the
quadrupole weight is spread over a finite interval, the poloidally induced
strain at $2\nu_R$ can remain larger than the local toroidal strain in any
individual bin, even when the integrated toroidal quadrupole budget is
larger.  This occurs because the toroidal contribution is distributed over
the interval $2\nu_R \le f_{\rm GW}\le 2(\nu_R+\Delta\nu)$, whereas the
poloidal contribution remains concentrated near $2\nu_R$.

The smearing is observationally resolved only when the bandwidth exceeds
the Fourier resolution, i.e., when
\begin{equation} \label{eq:resolved_smearing}
    \Delta f_{\rm GW}T \sim 2\Delta\nu T \gg 1 .
\end{equation}
If $2\Delta\nu T\lesssim1$, the signal is not resolved as a band and the
spectrum approaches the single-line limit.  When the condition
\eqref{eq:resolved_smearing} is satisfied, the detailed shape of the
smeared spectrum is controlled by $W_{ij}(\nu_Q)$ rather than by
$\Delta\nu/\nu_R$ alone.

\begin{figure}
\includegraphics[width=0.47\textwidth]{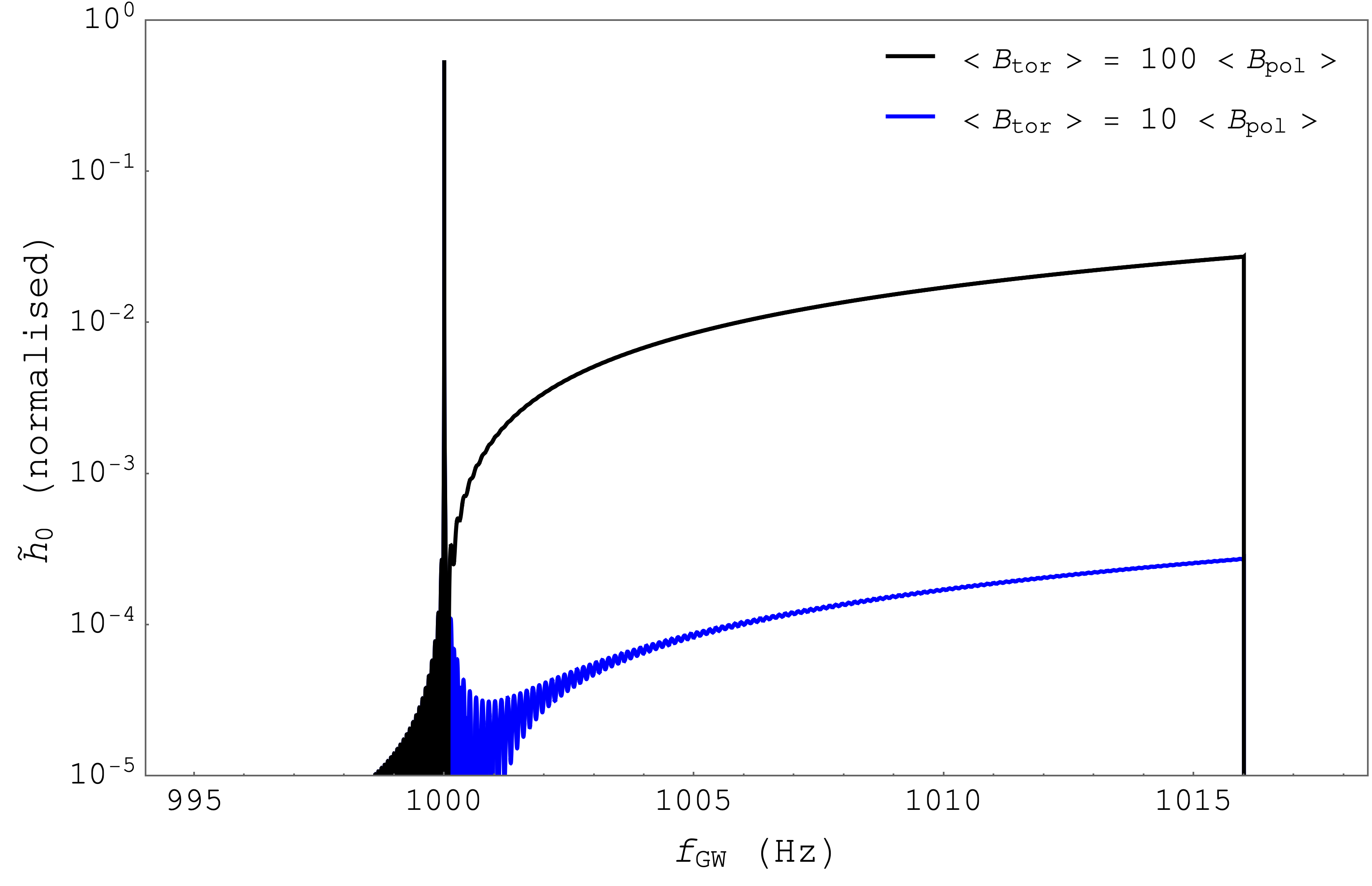}
\caption{Similar to Fig.~\ref{fig:magsmears} but for a square-root
   pattern-speed profile. As in Fig.~\ref{fig:magsmears},
  the vertical scale is normalised for display of the spectral
  structure  rather than for physical per-bin Fourier amplitudes; Eq.~\eqref{eq:perbin_reduction} gives the corresponding local amplitude scaling.}
\label{fig:magsmears2}
\end{figure}

To demonstrate a non-uniform toy spectrum, a second case, where instead
we posit a $\nu_\Delta$ that varies as the square root of distance from
toroidal edges to their epicentres, is depicted in
Figure~\ref{fig:magsmears2}.  Such a choice is motivated by the fact that
integrals of $\text{sinc}[T/2 (\omega - 2 \nR - 2 \Delta \nu)]$ can be
evaluated in terms of standard special functions if
$\Delta \nu \propto x^{1/p}$, where $x$ is the distance from the
boundary layer for integer $p$, making for a straightforward
evaluation (effectively with $p=1,2$ in Figs. \ref{fig:magsmears} and
\ref{fig:magsmears2}, respectively).  In this case, the shallower
gradient of $\nu_\Delta$ places more quadrupole weight near the upper
edge of the band, so the toroidal contribution peaks closer to
$2\nu_R+2\Delta\nu$ rather than being uniformly distributed.  This
illustrates a key feature of the smearing: a large toroidal
quadrupole can still yield a small local Fourier amplitude in each bin if
its weight is distributed over a wide frequency band.

While we emphasise again that we have not determined $\nD$
self-consistently here, the qualitative trends are robust and
demonstrate that GW power need not be concentrated at any given peak
if the mass quadrupole moment is induced by the magnetic field and
parts of the star are magnetically unlocked.  Even if the true
relaxation timescale is shorter than the estimate in
Eq.~\eqref{eq:tES_beta_corrected}, the qualitative effect remains: a
non-uniform pattern speed spreads the toroidal quadrupole signal over
a finite bandwidth. The detectability then depends on whether the
search strategy integrates power over that full band, rather than only
near $2\nR$.

\section{Impact on source detectability} \label{sec:detectability}

Having established that quadrupole weight can be redistributed outside the
conventional bin for isolated pulsars (as in Figs.~\ref{fig:magsmears} and \ref{fig:magsmears2}), it is
worth examining what this implies in light of current strain limits
\cite[see, e.g.,][]{wette23}.

Suppose first that a coherent search is conducted using a single matched
filter under the assumption that the signal is monochromatic over the
observing time.  If the quadrupole weight is instead distributed over a
bandwidth $\Delta f_{\rm GW}$, the band contains approximately
\begin{equation}
    N_{\rm bin}\sim \Delta f_{\rm GW}T
\end{equation}
independent frequency bins.  A monochromatic template then constrains only
the quadrupole weight lying inside the bin to which it is matched.  For a
uniform distribution with fixed total quadrupole weight, the coherent
Fourier amplitude in one bin is suppressed relative to the corresponding
line by $|\tilde h_{\rm bin}|/|\tilde h_{\rm line}| \sim N_{\rm bin}^{-1}
    \sim (\Delta f_{\rm GW}T)^{-1}$ (see equation \ref{eq:perbin_reduction}).
This is the scaling relevant for the local Fourier amplitudes displayed in
the smeared spectra.

This local coherent-amplitude suppression should not be confused with the
loss factor obtained when a search combines power over the whole smeared
band.  If the noise were approximately flat and the distribution known, a
band-combined or semi-coherent statistic could recover part of the lost
sensitivity by summing power across the $N_{\rm bin}$ bins.  The resulting
constraint on the total frequency-integrated quadrupole budget would then
be weakened by a factor of order
\begin{equation}
    N_{\rm bin}^{1/2}
    \sim (\Delta f_{\rm GW}T)^{1/2},
\end{equation}
or, equivalently, the corresponding amplitude sensitivity scales as
$(\Delta f_{\rm GW}T)^{-1/2}$.  Thus the per-bin coherent Fourier
amplitude scales as $(\Delta f_{\rm GW}T)^{-1}$, whereas a band-combined
constraint on the total quadrupole budget scales only as
$(\Delta f_{\rm GW}T)^{-1/2}$.
For
$\Delta\nu\sim1~{\rm Hz}$ and an observation time $T=1$ month,
$(\Delta f_{\rm GW}T)^{1/2}$ is of order $10^3$.  Even if the lag is much
more modest, say $\Delta\nu\sim1$ mHz, one still obtains a factor of order
$10^2$ for $T\sim1$ yr.  Tighter constraints on the total non-axisymmetric
quadrupole budget would therefore require search strategies that combine
power across the relevant band, such as cross-correlation or semi-coherent
methods adapted to a model for the quadrupole-weighted frequency
distribution.  In practice the noise spectral density $S_n(f)$ varies
across the band, so the optimal statistic must include the appropriate
frequency weighting, especially for non-uniform smearing \citep{jan98}.
Algorithms developed for spin wandering may also be useful if the
distribution evolves in time \citep{suva16}.

\section{Conclusions} \label{sec:conclusions}

In this paper, we reexamine the theory behind torque lags in neutron stars and argue that some differential rotation may persist in magnetically-unlocked regions for a physically plausible, though microphysics-dependent, range of parameters.
The main result is a mapping rather than a microphysical
prediction: if a non-axisymmetric quadrupole component is carried by a
magnetically-unlocked region, its radiation appears at twice the corresponding pattern frequency, or is redistributed over a band if the pattern speed is
spatially distributed.  Whether a large lag actually persists is
dependent on magnetic locking, viscous and thermal
transport, mutual friction, stratification, pairing, and the age and spin
history of the star.  In particular, the MSP estimates in
Sec.~\ref{sec:msps} and Appendix~\ref{sec:appendix} show that quantitative predictions for spectral shifts require not only an estimate of the internal temperature and a measured $\dot{\nu}_{\rm R}$, but a host of other assumptions including the quadrupole-weighted pattern-speed distribution.

For a uniform quadrupole-weighted distribution over a lag interval
$\Delta\nu$, the local coherent amplitude in one resolved Fourier bin is
suppressed relative to an unsmeared line by
$(Q_\Delta/Q_{\rm pol})(2\Delta\nu T)^{-1}$.  For the large MSP
baseline example
used to display the structure, $2T=1$ month, $\Delta\nu=7.9$ Hz, and
$Q_\Delta/Q_{\rm pol}=10^2$ give a local per-bin ratio of order
$\sim 10^{-5}$.  This per-bin scaling is distinct from the loss in a
band-combined constraint on the total frequency-integrated quadrupole
budget, which scales only as $(\Delta f_{\rm GW}T)^{1/2}$ in amplitude
sensitivity.  Thus spectral smearing redistributes the signal power over
frequency and, even for large quadrupoles, can render the signal in any given bin small (see Fig.~\ref{fig:magsmears2}).

The practical implication is therefore not that a lack of detections is
itself unsurprising, but that signals with redistributed
quadrupole weight need not be excluded by monochromatic-template limits
centred only at $2\nu_R$ or its harmonics.  Such limits constrain the quadrupole weight
inside the searched frequency window and must be interpreted together with a
model for $W_{ij}(\nu_Q)$.  The simple models used here should be refined in
future by self-consistently solving for differentially rotating, magnetised
equilibria \cite[see][]{uryu23}, but they already show why continuous-GW
searches that combine power across physically motivated frequency bands are
valuable.
Nevertheless, large ellipticities are not necessarily ruled out by a lack of GW detections
if the power is spread over a wide range, as can occur when the
effective secular coupling time is long.

\begin{acknowledgements}
We thank Kostas Glampedakis and Ian
Jones for valuable feedback.  A.~G.~S.  received funding from the
Deutsche Forschungsgemeinschaft via research grant 570901071 and the
European Union's Horizon MSCA-2022 research and innovation programme
``EinsteinWaves'' under grant agreement No. 101131233. A.~S. received
funding from the Polish NCN Grant No. 2023/51/B/ST9/02798 and in part by Deutsche
Forschungsgemeinschaft Grant No.  SE 1836/6-1.
\end{acknowledgements}

\bibliographystyle{aa}
\bibliography{magnetarbib}

\appendix

\begin{appendix}

\section{Ekman pumping, thermal and viscous timescales} \label{sec:appendix}

This appendix collects the order-of-magnitude timescales used in
Sec.~\ref{sec:timescales}.  The first two estimates, the Ekman time
and the zero-field shear-viscous diffusion time, are useful reference
values.  They are not, however, always the appropriate relaxation
times for a magnetically-unlocked toroidal region.  In such a region
the angular-momentum flux required to erase a lag is transverse to the
local magnetic surfaces, so the relevant viscosity is a component of
the magnetised viscosity tensor.  Moreover, if the neutrons are
paired, the neutron quasiparticle contribution can be strongly
suppressed.  We therefore distinguish below between the isotropic
reference quantity $\tau_\eta$ and the transverse magnetised estimate $\tau_{\eta,\perp}$.

For the leptonic transport coefficients we quote normal-proton
expressions as a limiting-case example.  In realistic mature neutron
stars, protons may be superconducting, while sufficiently strong
internal magnetar fields can locally suppress proton pairing.  The
precise transport coefficients therefore depend on the superconducting
state and field strength.  The aim here is to obtain simple estimates
of timescales on which the closed toroidal magnetic region couples to
the external electromagnetic torque.

\subsection{Ekman pumping}

In an unstratified Newtonian fluid, changes in the crustal rotation
rate are communicated to the interior through Ekman pumping.  The
Coriolis force drives a secondary circulation within boundary layers
of thickness
\begin{equation}
\delta_E \sim E^{1/2}R,\quad E=\frac{\nu}{R^2\Omega},
\end{equation}
where $E$ is the Ekman number and $\nu$ is the kinematic viscosity (not to be confused with the rigid rotation rate $\nR$). 

Fluid within the boundary layer is transported radially and recycled
through the stellar interior, spinning down the bulk fluid on the
Ekman timescale
\begin{equation}
\tau_E^{-1} = E^{1/2}\Omega,
\end{equation}
which on substituting $E$ translates to $\tau_E \sim R/\!\sqrt{\nu\Omega}$.

In a neutron-star core with superfluid neutrons and normal protons,
the shear viscosity is determined primarily by leptons, with the
dominant contribution coming from electron--electron collisions
\citep{Flowers1979,Shternin2008}.  Adopting the $e$--$e$ viscosity for
ultrarelativistic degenerate electrons in the
transverse-plasmon/Landau-damping regime from \cite{Shternin2008}:
\begin{equation}\label{eq:eta_ee}
\eta_{ee} \simeq 8.6\times 10^{18} \left(\frac{Y_e}{0.05}\right)^{14/9} \left(\frac{\rho}{\rho_0}\right)^{14/9} T_8^{-5/3} \mathrm{g\,cm^{-1}\,s^{-1}}, 
\end{equation}
where $T_8=T/10^8\ \mathrm{K}$.  The corresponding kinematic
viscosity $\nu_{ee}=\eta_{ee}/\rho$ is
\begin{equation}
\nu_{ee}
\simeq
3.1\times10^4
\left(\frac{Y_e}{0.05}\right)^{14/9}
\left(\frac{\rho}{\rho_0}\right)^{5/9}
T_8^{-5/3}
\ \mathrm{cm^2\,s^{-1}},
\end{equation}
where $Y_e$ is the electron fraction.
Substituting into the definition of the Ekman number gives
\bea\label{eq:Ekman_number}
E_{ee}&\simeq&
3.1\times10^{-10}
\left(\frac{Y_e}{0.05}\right)^{14/9}
\left(\frac{\rho}{\rho_0}\right)^{5/9}
T_8^{-5/3}
\nonumber\\
&\times&
\left(\frac{R}{10^6\ \mathrm{cm}}\right)^{-2}
\left(\frac{\Omega}{100\ \mathrm{s}^{-1}}\right)^{-1}.
\eea
This is close to the estimate of \cite{mel12}, with the main
differences arising from our choices of composition and density
scaling.  The corresponding Ekman time is
\bea
\tau_E\simeq
570
\left(\frac{E_{ee}}{3.1\times10^{-10}}\right)^{-1/2}
\left(\frac{\Omega}{100\ \mathrm{s}^{-1}}\right)^{-1}  \mathrm{ s}.
\eea
Thus, the electron-viscosity Ekman coupling timescale is
$\tau_E\sim10^2$--$10^3$~s for MSPs
($\Omega\sim10^2$--$10^3$~s$^{-1}$), and $\tau_E\sim$ hours to days for
slowly rotating magnetars ($\Omega\sim1$~s$^{-1}$).

If Ekman pumping operates without impediment, differential rotation is
erased on a timescale of order $\tau_E$.  However, two distinct physical mechanisms
can suppress the Ekman secondary circulation in a
strongly stratified superfluid core: (i)~stable compositional
stratification provides a buoyancy force that inhibits radial fluid
displacements, and (ii)~mutual friction between superfluid vortices
and the normal-fluid component damps the secondary flow.  Either
mechanism can render $\tau_E$ inapplicable as the actual relaxation
timescale of the differentially-rotating region.

\subsection{Shear-viscous diffusion time and magnetised transverse viscosity}

Differential rotation can be smoothed directly by shear viscosity.  In an unmagnetised isotropic fluid, the corresponding diffusion time is
\begin{equation}\label{eq:teta_general}
 \tau_\eta \equiv \frac{R^2}{\nu_\eta}
 =\frac{\rho R^2}{\eta}
 =\frac{1}{E_\eta\,\Omega},
\end{equation}
where $\nu_\eta=\eta/\rho$ and $E_\eta=\nu_\eta/(\Omega R^2)$.
Adopting Eq. \eqref{eq:eta_ee} 
one obtains 
\bea\label{eq:eta_ee_iso}
\tau_\eta^{(ee)}&\simeq{}&1.02\,
\left(\frac{Y_e}{0.05}\right)^{-14/9}
\left(\frac{\rho}{\rho_0}\right)^{-5/9}
T_8^{5/3}
\left(\frac{R}{10^6\,\mathrm{cm}}\right)^2
\ {\rm yr}.
\eea
This is the familiar result that implies that the zero-field electron
viscosity smooths an unstratified velocity perturbation on a timescale
of order one year.  It should not be applied without modification to
angular-momentum transport across a strong toroidal magnetic field.

Let the magnetic field be locally along the azimuthal direction, as in a
toroidal zone, while the gradient of the differential rotation is normal to
the magnetic surface. The relevant viscous stress then involves a transverse
component of the viscosity tensor. In the relaxation-time approximation this
may be estimated as \citep{ShterninOfengeim2022}
\bea
 \eta_{e,\perp}\simeq \frac{\eta_{ee}}{1+x_H^2},
 \qquad
 x_H=\omega_{Be}\tau_{e,{\rm tr}}^{(\eta)} .
\eea
Here $x_H$ is the electron magnetisation parameter, and
$\tau_{e,{\rm tr}}^{(\eta)}$ is the microscopic electron transport
relaxation time associated with shear viscosity. It should not be confused
with the macroscopic viscous diffusion time $\tau_\eta=\rho R^2/\eta$
defined above.

For ultrarelativistic degenerate electrons,
\bea
 \eta_{ee}\simeq
 \frac{1}{5}n_e p_{Fe}c\,\tau_{e,{\rm tr}}^{(\eta)}
 =
 \frac{1}{5}n_e p_{Fe}\lambda_{e,{\rm tr}}^{(\eta)},
 \qquad
 \lambda_{e,{\rm tr}}^{(\eta)}=c\tau_{e,{\rm tr}}^{(\eta)} .
\eea
Therefore,
\bea
 \tau_{e,{\rm tr}}^{(\eta)}
 =
 \frac{5\eta_{ee}}{n_e p_{Fe}c},
\eea
Using Eq.~\eqref{eq:eta_ee}, with
$n_e=Y_e\rho/m_b$, where $m_b$ is the baryon mass, and
$p_{Fe}=\hbar(3\pi^2 n_e)^{1/3}$, gives
\bea
 \tau_{e,{\rm tr}}^{(\eta)}
 \simeq
 2.56\times10^{-14}
 \left(\frac{Y_e}{0.05}\right)^{2/9}
 \left(\frac{\rho}{\rho_0}\right)^{2/9}
 T_8^{-5/3}\ {\rm s}.
\eea
For relativistic electrons,
\bea
 \omega_{Be}\simeq \frac{eB}{p_{Fe}},
\eea
and hence
\bea
 x_H\simeq 1.85
 \left(\frac{B}{10^9\,{\rm G}}\right)
 T_8^{-5/3}
 \left(\frac{Y_e}{0.05}\right)^{-1/9}
 \left(\frac{\rho}{\rho_0}\right)^{-1/9}.
\eea
For the fiducial values $\rho=\rho_0$ and $Y_e=0.05$, this reduces to
\bea
 x_H\simeq 1.85
 \left(\frac{B}{10^9\,{\rm G}}\right)
 T_8^{-5/3}.
\eea
The electron contribution to the transverse viscous diffusion time across a
region of size $L_\Delta$ is therefore
\bea\label{eq:tau_eta_eperp}
 \tau_{\eta,e\perp}
 \simeq
 1.49
 \left(\frac{L_\Delta}{R}\right)^2
 T_8^{5/3}
 \left[
 1+3.43
 \left(\frac{B}{10^9\,{\rm G}}\right)^2
 T_8^{-10/3}
 \right]
 \ {\rm yr},
\eea
where $R=12$ km, $\rho=\rho_0$, and $Y_e=0.05$ have been used in the
numerical coefficient. For $x_H\gg1$, the transverse electron viscosity is
strongly suppressed and the corresponding macroscopic diffusion time
$\tau_{\eta,e\perp}$ can become very long.

The total transverse viscosity also contains a neutron quasiparticle
contribution in the superfluid phase $\eta_n^{\rm sf}$ .  We write
\begin{equation}\label{eq:eta_eff_perp}
 \eta_{\rm eff,\perp}=\eta_{e,\perp}+\eta_n^{\rm sf},
 \qquad
 \tau_{\eta,\perp}=\frac{\rho L_\Delta^2}{\eta_{\rm eff,\perp}}.
\end{equation}

For the normal neutron viscosity we use the standard estimate
\begin{equation}\label{eq:eta_n_normal}
\eta_n^N\simeq
1.15\times10^{19}
\left(\frac{\rho}{\rho_0}\right)^{9/4}
T_8^{-2}
\ {\rm g\,cm^{-1}\,s^{-1}},
\end{equation}
and suppress it by neutron pairing according to
\begin{equation}\label{eq:eta_n_sf}
 \eta_n^{\rm sf}=R_n(y)\eta_n^{N},
 \qquad y=\frac{\Delta_n}{k_BT},
\end{equation}
with the fit \citep{Andersson04},
\begin{eqnarray}
 R_n(y)&=&
 \left[0.9543+\sqrt{0.04569^2+(0.6971y)^2}\right]^3
 \nonumber\\
 &&\times
 \exp\left[0.1148-\sqrt{0.1148^2+4y^2}\right].
\end{eqnarray}
The size of this suppression is very sensitive to the
${}^3P_2$--${}^3F_2$ gap.  Microscopic calculations summarised by
\cite{SedrakianClark2019} give gaps of several tenths of an MeV around
$k_{Fn}\simeq1.7~{\rm fm}^{-1}$, although smaller gaps are possible
when medium effects are included.

For the magnetar example  used in the main text, $T=6\times10^8$ K and $B_{\rm tor}=10^{14}$ G, the estimated relaxation times are
\begin{equation}
\begin{array}{c|ccc}
\Delta_n & \tau_{\eta,\perp}(0.1R) & \tau_{\eta,\perp}(0.3R) &
                                                               \tau_{\eta,\perp}(R)
                                                               \\
\hline
0.2~{\rm MeV} & 1.7\times10^{1} & 1.5\times10^{2} & 1.7\times10^{3}\\
0.4~{\rm MeV} & 7.4\times10^{3} & 6.6\times10^{4} & 7.4\times10^{5}\\
0.6~{\rm MeV} & 4.8\times10^{6} & 4.3\times10^{7} & 4.8\times10^{8}
\end{array}
\quad {\rm yr}.
\label{eq:visc_table}
\end{equation}
Thus, for $\Delta_n\simeq0.6$ MeV and a toroidal zone of size $L_\Delta\gtrsim0.1R$, transverse viscous diffusion is slower than the composition-controlled time \eqref{eq:tES_beta_corrected}; for $\Delta_n\simeq0.4$ MeV the two are comparable for stellar-scale regions; and for $\Delta_n\simeq0.2$ MeV viscosity remains the faster secular relaxation channel.

\subsection{Eddington--Sweet relaxation for normal protons}

To maintain consistency with the normal-proton evaluation of the
electron shear viscosity and the Ekman number, the thermal conductivity
must also be evaluated using the normal-state transverse photon
propagator.  In this regime the transverse electromagnetic interaction
is dynamically screened by Landau damping rather than by the Meissner
response of a proton condensate.

For normal protons, \cite{Shternin2007} obtain for each
lepton species $\ell=e,\mu$
\bea
\kappa_\ell^{(N)}
=
\frac{\pi^2}{54\zeta(3)}
\frac{k_{\rm B}c}{\alpha}
\left(\frac{p_{F\ell}}{\hbar}\right)^2 ,
\label{eq:kappa_normal_lepton}
\eea where the symbols have their usual meaning (e.g., $\alpha$ is the
fine structure constant).  The conductivity is approximately
temperature independent, in contrast to the usual Fermi-liquid
behaviour $\kappa\propto T^{-1}$ that applies in the
superconducting-proton case.  The total lepton conductivity is \bea
\kappa_{e\mu}^{(N)} = \kappa_e^{(N)}+\kappa_\mu^{(N)}.  \eea Using the
relation $p_{Fe}=\hbar(3\pi^2n_e)^{1/3}$, the electron contribution
can be written in the fiducial form
\bea \kappa_e^{(N)}
\simeq 3.4\times10^{21} \left(\frac{Y_e}{0.05}\right)^{2/3}
\left(\frac{\rho}{\rho_0}\right)^{2/3} \
\frac{\mathrm{erg}}{\mathrm{cm\,s\,K}}.
\label{eq:kappa_e_normal_fid}
\eea
Including muons amounts to multiplying
Eq.~\eqref{eq:kappa_e_normal_fid} by the factor
\bea
\Phi_\mu
=
1+
\left(\frac{n_\mu}{n_e}\right)^{2/3}.
\eea
In terms of thermal diffusivity 
$\chi=\kappa/C_V,$
the Eddington--Sweet relaxation time is given by
\bea
\label{eq:tkappa_general}
\tau_\kappa
= S\,\frac{R^2}{\chi}
= S\,\frac{R^2 C_V}{\kappa}.
\eea
For the stratification parameter $S$, we use equation~\eqref{eq:S}.  If the
neutrons are superfluid but the protons are normal, the heat capacity
receives contributions from both leptons and protons (the neutron
quasiparticle contribution is exponentially suppressed by the pairing
gap).  For ultrarelativistic electrons,
\begin{equation}
C_{V,e}
\simeq
7.9\times10^{17}
\left(\frac{Y_e}{0.05}\right)^{2/3}
\left(\frac{\rho}{\rho_0}\right)^{2/3}
T_8
 \frac{\mathrm{erg}}{\mathrm{cm^3\,K}},
\end{equation}
 whereas for degenerate nonrelativistic protons,
\bea
C_{V,p}
\simeq
4.2\times10^{18}
\left(\frac{m_p^\ast}{0.7m_p}\right)
\left(\frac{Y_p}{0.05}\right)^{1/3}
\left(\frac{\rho}{\rho_0}\right)^{1/3}
T_8
\ \frac{\mathrm{erg}}{\mathrm{cm^3\,K}}.
\eea
 Neglecting muons for simplicity,
\bea
C_V
\simeq
5.0\times10^{18}\,T_8
\ \frac{\mathrm{erg}}{\mathrm{cm^3\,K}},
\eea
where the proton term is dominant but the electron contribution has been
included in the numerical coefficient. 
The corresponding thermal diffusivity is
\bea
\chi_{e}^{(N)}
&\simeq&
6.8\times10^2
\left(\frac{m_p^\ast}{0.7m_p}\right)^{-1}
\left(\frac{Y_p}{0.05}\right)^{-1/3}\nonumber\\
&\times&\left(\frac{Y_e}{0.05}\right)^{2/3}
\left(\frac{\rho}{\rho_0}\right)^{1/3}
T_8^{-1}
\ \mathrm{cm^2\,s^{-1}},
\eea
where the proton contribution to $C_V$ has been retained.
Substituting the above into Eq.~\eqref{eq:tkappa_general} gives
\bea
\label{eq:tkappa_normal}
\tau_\kappa^{(N)}
&\simeq&
1.35\times10^4\ \mathrm{yr}\,
\left(\frac{R}{10^6\,\mathrm{cm}}\right)^2
\left(\frac{\Omega}{100\,\mathrm{s}^{-1}}\right)^{-2}
T_8
\nonumber\\
&\times&
\left(\frac{m_p^\ast}{0.7m_p}\right)
\left(\frac{Y_p}{0.05}\right)^{1/3}
\left(\frac{Y_e}{0.05}\right)^{-2/3}
\Phi_\mu^{-1}.
\eea
For charge-neutral $npe$ matter, $Y_p=Y_e$, and the residual
composition dependence becomes weak and we have
\bea
\tau_\kappa^{(N)}
&\simeq&
1.35\times10^4\ \mathrm{yr}\,
\left(\frac{R}{10^6\,\mathrm{cm}}\right)^2\nonumber\\
&\times&\left(\frac{\Omega}{100\,\mathrm{s}^{-1}}\right)^{-2}
T_8
\left(\frac{Y_e}{0.05}\right)^{-1/3}.
\eea
If the proton heat capacity is omitted and only the leptonic heat
capacity is retained, the normal-proton estimate reduces to
\begin{equation} \label{eq:tkappa_normal_leptons}
\tau_{\kappa,\ell}^{(N)}
\simeq
2.1\times10^3\ \mathrm{yr}\,
\left(\frac{R}{10^6\,\mathrm{cm}}\right)^2
\left(\frac{\Omega}{100\,\mathrm{s}^{-1}}\right)^{-2}
T_8.
\end{equation}
This expression is not a superconducting-proton result; it is only the
lepton-heat-capacity limit of the normal-proton calculation.  The larger
value in Eq.~\eqref{eq:tkappa_normal} is the consistent choice when
protons are normal, because normal protons contribute substantially to
the heat capacity even though the heat {current} is carried primarily
by leptons.

An additional estimate concerns the neutron thermal conductivity.
If the neutrons are normal, their conductivity generally exceeds the
lepton conductivity at $T\sim10^8$~K, and the total conductivity should
be written as $\kappa_{\rm tot}=\kappa_n+\kappa_{e\mu}$.  In that case
the actual $\tau_\kappa$ is shorter than the lepton-only estimate above.
If the neutrons are strongly superfluid, their quasiparticle conductivity
and heat capacity are both suppressed by the pairing gap, and
Eq.~\eqref{eq:tkappa_normal} provides the appropriate normal-proton,
lepton-transport estimate.

The temperature dependence in Eq.~\eqref{eq:tkappa_normal} follows from
the normal-proton conductivity.  Since $\kappa_{e\mu}^{(N)}$ is nearly
temperature independent while $C_V\propto T$, one obtains
$\tau_\kappa^{(N)}\propto T$.  In a strongly superconducting-proton
regime the transverse photon propagator is modified by the Meissner
response, and the lepton thermal conductivity has a different temperature
dependence.  In that case one expects $\kappa_{e\mu}\propto T^{-1}$ and
hence $\tau_\kappa\propto T^2$, but a self-consistent
superconducting-proton transport calculation is beyond the scope of the
present discussion aimed at order-of-magnitude estimates.

\subsection{Comparison of timescales}

The estimates above clarify how the timescale hierarchy should be interpreted.  The Ekman time,
\begin{equation}
 \tau_E\sim10^2\text{--}10^3~{\rm s},
\end{equation}
for rapidly rotating objects, describes the rapid boundary-layer response in an unstratified or weakly stratified fluid.  In a strongly stratified core it need not be the time for global corotation.  The zero-field isotropic viscous time,
\begin{equation}
 \tau_\eta^{(ee)}\sim1~{\rm yr},
\end{equation}
is likewise a useful baseline, but it is not necessarily the appropriate torque time for a closed toroidal magnetic region.

The relevant secular comparison is instead between complete channels,
\begin{equation}
 \frac{1}{\tau_{\rm eff}}
 \simeq
 \frac{1}{\tau_B}
 +\frac{1}{\tau_{\eta,\perp}}
 +\frac{1}{\tau_\kappa}
 +\frac{1}{\tau_{\rm ES}^{(\beta)}}
 +\frac{1}{\tau_{\rm mf}}
 +\cdots .
\end{equation}
If magnetic stresses connect the region to the crust, $\tau_B$ is short and the lag is quickly removed.  If the region lies on closed field lines or in a magnetic void, this channel is strongly reduced.  If thermal diffusion drives a complete meridional-circulation response, the relevant time is $\tau_\kappa$ from Eq.~\eqref{eq:tkappa_general}.  If mutual friction is efficient, $\tau_{\rm mf}$ controls the coupling to the neutron superfluid; if it is weak, the superfluid can remain differentially rotating on secular timescales \citep{Sedrakian1998}.  The composition-controlled Eddington--Sweet time, $\tau_{\rm ES}^{(\beta)}$, is the timescale for a beta-limited meridional-circulation channel and should not be combined with $\tau_\beta$ as a separate inverse rate.

For a magnetar-like example with $T=6\times10^8$ K
and $\nu_R=0.1$ Hz, Eq.~\eqref{eq:tES_beta_corrected} gives
\begin{equation}
 \tau_{\rm ES}^{(\beta)}\simeq8\times10^5~{\rm yr}.
\end{equation}
Equation~\eqref{eq:visc_table} shows that the direct transverse
viscous channel may be faster or slower than this value depending
primarily on the neutron triplet gap and on the size of the toroidal
region.  A gap $\Delta_n\simeq0.2$ MeV leaves a large neutron viscous
floor and leads to relatively rapid relaxation.  A gap
$\Delta_n\simeq0.6$ MeV suppresses the neutron contribution strongly
enough that $\tau_{\eta,\perp}$ can exceed $\tau_{\rm ES}^{(\beta)}$
for plausible magnetar-strength toroidal fields and kilometre-scale
regions.  The intermediate value $\Delta_n\simeq0.4$ MeV is
transitional and depends on microscopic and geometric uncertainties.

For MSPs, the hierarchy is different.  At
$\nu_R=500$~Hz the stiffness parameter is only $S\simeq0.3$, so the
strong-stratification argument is much weaker.  The normal-proton
thermal estimate gives $\tau_\kappa^{(N)}\sim$ a few years for
$T=3\times10^7$~K, while the transverse electron-viscous time ranges
from tens of years to $10^5$--$10^6$~yr depending on $B_{\rm tor}$ and
$L_\Delta$.  These channels are faster than
$\tau_{\rm ES}^{(\beta)}$ unless they are geometrically or
microphysically suppressed.  Therefore the beta-limited lag used in the
MSP spectral examples is an extreme upper-bound limiting case. 

To conclude, long-lived differential rotation is not guaranteed by
compositional stratification alone.  It is possible when rapid
magnetic locking is absent and when the remaining channels (transverse
viscosity, thermal circulation, mutual friction, and beta-limited
circulation) are all secularly slow.  In the opposite regime,
viscosity, thermal circulation, or mutual friction or their
combination wins and the lag is correspondingly reduced.  The spectral
calculations in the main text should be understood as a mapping from a
given quadrupole-weighted distribution of pattern speeds to the GW
spectrum, with the distribution itself strongly dependent on the
uncertain microphysics.
\end{appendix}

\end{document}